\documentclass[pra,showpacs,preprintnumbers,amsmath,amssymb,tightenlines,superscriptaddress]{revtex4}

\usepackage{graphicx}
\usepackage{dcolumn}
\usepackage{bm}
\usepackage{epsfig}
\usepackage{stmaryrd}
\usepackage{dsfont}
\usepackage{amssymb}
\usepackage{bm}
\usepackage{color}

\def\balpha{\bm{\alpha}}
\def\bbeta{\bm{\beta}}
\def\bgamma{\bm{\gamma}}
\newcommand{\aop}[1]{\hat{a}_{#1}}
\newcommand{\adag}[1]{\hat{a}^{\dagger}_{#1}}

\newcommand{\ssection}[1]{{\noi  \it #1:}}

\newcommand{\ket}[1]{|\,{#1}\,\rangle}

 \newcommand{\pdiff}[2]{\frac{\partial #1}{\partial #2}}

\newcommand{\expec}[1]{\langle #1 \rangle}

\newcommand{\movel}{\!\!\!\!\!\!\!\!\!\!\!\!\!\!\!\!\!\!\!}

\newcommand{\sub}[2]{{#1}_{ \mbox{\scriptsize #2}}}
\newcommand{\mv}[1]{\mathbf{#1 }}
\newcommand{\bv}[1]{\bm{ #1 }}
\newcommand{\bvv}[1]{\underline{\bm{ #1 }}}
\newcommand{\mvv}[1]{\underline{\underline{ #1 }}\,}

\newcommand{\id}{\mathds{1}}

\newcommand{\psiop}[1]{\hat{\Psi}_{#1}(\mv{x})}
\newcommand{\psidag}[1]{\hat{\Psi}^{\dagger}_{#1}(\mv{x})}

\def\noi{\noindent}
\def\beq{\begin{equation}}
\def\eeq{\end{equation}}

\def\CR{\nonumber\\[0.15cm]}

\newcommand{\fref}[1]{Fig.~\ref{#1}}
\newcommand{\frefp}[2]{Fig.~\ref{#1}~(#2)}

\newcommand{\eref}[1]{Eq.~(\ref{#1})}

\newcommand{\sref}[1]{section~\ref{#1}}
\newcommand{\cref}[1]{chapter~\ref{#1}}
\newcommand{\tref}[1]{table~\ref{#1}}
\newcommand{\aref}[1]{\ref{#1}}
\newcommand{\bref}[1]{(\ref{#1})}

\newcommand{\wt}[1]{\widetilde{#1}}

\begin{document}

\title{Quantum dynamics of long-range interacting systems using the positive-P and gauge-P representations}
\author{S.~W\"uster}
\affiliation{Max Planck Institute for the Physics of Complex Systems, N\"othnitzer Strasse 38, 01187 Dresden, Germany}
\affiliation{Department of Physics, Bilkent University, Ankara 06800, Turkey}
\affiliation{Department of Physics, Indian Institute of Science Education and Research, Bhopal, Madhya Pradesh 462 023, India}
\author{J.~F.~Corney}
\affiliation{School of Mathematics and Physics, University of Queensland, Brisbane QLD 4072, Australia}
\author{J.~M.~Rost}
\affiliation{Max Planck Institute for the Physics of Complex Systems, N\"othnitzer Strasse 38, 01187 Dresden, Germany}
\author{P.~Deuar}
\affiliation{Institute of Physics, Polish Academy of Sciences, Aleja Lotnik\'ow 32/46, Pl-02-668 Warsaw, Poland}
\email{sebastian@iiserb.ac.in}
\begin{abstract}
We provide the necessary framework for carrying out stochastic positive-P and gauge-P simulations of bosonic systems with long range interactions. In these approaches, the quantum evolution is sampled by trajectories in phase space, allowing calculation of correlations without truncation of the Hilbert space or other approximations to the quantum state. The main drawback is that the simulation time is limited by noise arising from interactions.
 We show that the long-range character of these interactions does not further increase the limitations of these methods, in contrast to the situation for alternatives such as the density matrix renormalisation group. Furthermore, stochastic gauge techniques can also successfully extend simulation times in the long-range-interaction case, by making using of parameters that affect the noise properties of trajectories, without affecting physical observables.

We derive essential results that significantly aid the use of these methods: estimates of the available simulation time, optimized stochastic gauges, a general form of the characteristic stochastic variance and adaptations for very large systems. Testing the performance of particular {\it drift} and {\it diffusion} gauges for nonlocal interactions, we find that, for small to medium systems, drift gauges are beneficial, whereas for sufficiently large systems, it is optimal to use only a diffusion gauge.

The methods are illustrated with direct numerical simulations of interaction quenches in extended Bose-Hubbard lattice systems and the excitation of Rydberg states in a Bose-Einstein condensate, also without the need for the typical frozen gas approximation. We demonstrate that gauges can indeed lengthen the useful simulation time. 
\end{abstract}

\pacs{
2.50.Ey, 
03.75.Gg, 
05.10.G
}

\maketitle

\section{Introduction}

First-principles treatments of many-body quantum problems are notoriously difficult due to the exponential increase of Hilbert space dimension with the number of system components. Tackling this complexity is an outstanding goal of theoretical physics. 
Long-range interactions usually amplify the difficulties, because they can break symmetries or frustrate many of the tricks used to reduce the Hilbert space to manageable sizes. For example, the entanglement area laws can be broken \cite{Eisert10}.

If limited precision is acceptable, stochastic phase-space methods are a promising contender compared to other standard methods such as Monte Carlo or entangled pair states~\cite{steel:wigner,book:qn,drummond:review:jmo}, especially for higher-dimensional systems. The computational complexity of phase-space methods tends to scale only linearly or quadratically in system size and to be largely independent of dimensionality. Originating from quantum optics~\cite{drummond:gardiner:posp}, these methods have also been successful for cold degenerate gases~\cite{karen:epr,craig:dissoc,Carusotto01a,Carusotto01b,deuar:collisions,deuar:canonical,Polkovnikov10,stab,sqcoll,Kheruntsyan12,Lewis-Swan14,Lewis-Swan15} including fermions \cite{Corney04,Corney06} and spin systems \cite{Barry08,Ng13,anamaria1}. They are particularly suited to relatively dilute boson systems deep in the quantum regime. 

 Apart from a few small forays \cite{drummond:review:jmo,DeuarPhD-ranged}, previous work with phase-space methods has been limited to systems with contact interactions.
However, long-range interactions have recently become important in the dilute quantum regime due to advances in the production of ultracold Rydberg \cite{book:gallagher,loew:tutorial},  dipolar atomic \cite{Griesmaier:chromiumcondensate,Lahaye:dipolar:review} and molecular \cite{carr:ucoldmol:review} gases, whose physics can not be captured with a contact interaction alone.
Motivated by this experimental progress, we develop here tools for the application of phase-space methods to systems with long range interactions. These tools could then go substantially beyond the initial exploration of the positive-P method to a Rydberg system, reported in \cite{wu:laserdrivenryd:posp}, which required a simplified interaction model.

The price paid for use of phase-space methods is that the dynamics is tractable usually only over short time scales. This limitation arises from the nonlinear amplification of stochastic fluctuations in individual trajectories, leading to a phase-space distribution that is too broad~\cite{gardiner:boundaries,deuar:pra,deuar:jphysAI,Carusotto03d}. Hence, a central requirement in practice is to estimate the time limits and if possible to extend them. 
Fortunately, one can exploit the over-completeness of the basis used to keep the distribution as compact as possible and to stabilize trajectories. Simulation schemes exploiting this feature have been termed Gauge-P methods~\cite{plimak:diffusion,deuar:pra,dowling04,deuar:job,deuar:jphysAII, deuar:jcpc,deuar:thesis,deuar:canonical,deuar:collisions}. Two different types of gauges are available. \emph{Drift}-gauges modify the drift term of the underlying Fokker-Planck equation for the phase-space distribution, allowing e.g.~a modification of nonlinearities in the evolution equation at the expense of introducing a fluctuating trajectory weight. \emph{Diffusion} gauges alter the noise terms in the stochastic evolution equations and can be used to vary the relative fraction of noise affecting the phase or amplitude of complex variables.

These stochastic gauges can become particularly powerful when it is possible to determine the best choice of gauge for a given simulation. For problems that rely essentially on single bosonic modes, these best choices have been found~\cite{deuar:thesis,deuar:jphysAII}. For the many-mode case to date, optimization has only been considered insofar as a collection of single modes can offer some guidance, in situations where the interactions are weak or short range. 

In this work, we adapt the prescriptions and approaches that are successful for short range interactions to long-range interactions, and we show that they still give useful extensions of simulation time. 
To this end, we calculate the evolution of a characteristic variance that can serve to estimate the effect of a specific gauge, analogous to the approach of~\cite{deuar:jphysAII} for the single-mode case. Though quite an involved expression, it provides the essential basis for the application of Gauge-P techniques to problems with long-range interactions. Estimates of the available simulation time and optimal adaptive gauge choices then follow in a relatively straightforward way. 

The article is organized as follows: In \sref{manybody} we describe the class of many-body quantum mechanical models that are being considered here. The general idea behind the positive-P and Gauge-P methods is then briefly reviewed in \sref{method}, and the evolution equations for nonlocal interactions derived. In \sref{optimalgauges}, which contains the core of our analysis, we derive and analyze the characteristic noise variances and optimize gauges. Based on this, \sref{tsimsection} presents estimates of the available simulation time.

All these results are then applied to two test cases in \sref{MBsim}: an interaction quench in an extended (long-range) Bose-Hubbard model and the excitation dynamics of mobile Rydberg atoms within a BEC. 
These examples are oriented towards long-range interacting ultra-cold atomic physics, such as Bose-Einstein condensates (BEC) of dipolar atoms \cite{Griesmaier:chromiumcondensate}, polar molecules \cite{carr:ucoldmol:review} or Rydberg dressed condensates~\cite{santos:dressing,nils:supersolids,pupillo:dressed,honer:dressing,fabian:bullets,wuester:dressing,johnson:dressing,balewski:dressingNJP,jau:expt}. 
Following this, \sref{largeM} presents some further results that facilitate simulation of very large systems: a more efficient representation of the noise, and an optimization of pure diffusion gauges.
In \sref{nonlocal_diff}, we report also on some initial explorations of more general nonlocal gauges as a starting-point for future work.  
Appendices contain various technical details, such as the Stratonovich corrections (\aref{stratocorrapp}), the complete derivation of the characteristic variances (\aref{charvarapp} and \aref{charvarappP}), and supporting material on non-local noises (\aref{noiserewriting}) and gauges (\aref{nonlocalDG}).

\section{Long-range interacting systems}
\label{manybody}

The class of Hamiltonian operators with which we work in this article is represented by
\begin{eqnarray}
\hat{H}=\sum_{nm}\left[ \adag{n}\omega_{nm} \aop{m} + \frac{1}{2} \adag{n}\adag{m}W_{nm} \aop{n} \aop{m} \right].
\label{Hamiltonian}
\end{eqnarray}
The indices label elements of the chosen single particle basis $|m\rangle$ and the operator $\aop{m}$ creates a particle in that basis state. Single particle energies and linear mode couplings are included in $\omega_{nm}$, while $W_{nm}$ parametrizes generic two-body interactions. This potential $W$ is assumed to be real here. For Hermiticity, $\omega_{nm}=\omega_{mn}^*$.

We now introduce two physical model systems the Hamiltonian of which has the form \bref{Hamiltonian}. For these examples, we show simulation results in \sref{MBsim}.

\subsection{Extended Bose-Hubbard model}
\label{extendedBH_preview}

One Hamiltonian of the form \eref{Hamiltonian} is that of an extended (long-range interacting) Bose-Hubbard model. In a cold atom context, $|m\rangle$ is then the localized, lowest-band, Wannier function at a single site of an optical lattice. The coupling $\omega_{nm}=J(\delta_{n+1,m} + \delta_{n-1,m})$ describes particle tunneling between neighboring lattice sites with amplitude $J$. Many cases of long-range interactions can be captured by the potential
\begin{eqnarray}
W(|\mv{x}-\mv{y}|)=-\frac{C_{6}}{(|\mv{x}-\mv{y}|^{a} + \epsilon^{a})^b}.
\label{potential}
\end{eqnarray}
Possible experimental realizations include optical lattices filled with polar atoms \cite{Griesmaier:chromiumcondensate} or Rydberg dressed atoms \cite{laura:latticeclock,macri:latticedress,weibin:latticemotion:prl,zeiher:dressedlattice}. The $W_{nm}$ in \bref{Hamiltonian} can then be obtained from \eref{potential} by use of overlap integrals involving Wannier basis functions.  However, here we simply use $W_{nm}=W(r)$, where $r=|\mv{x}-\mv{y}|$ and $\mv{x}_{n}$, $\mv{x}_{m}$ are discrete lattice points. We will later present simulation results for $a=2$ and $b=3$, corresponding to a softened van-der-Waals potential. Alternatively we could have used $a=6$ and $b=1$ to represent Rydberg-dressed potentials that include the cut-off due to blockade effects \cite{nils:supersolids}. The contact part of the interaction plays a prominent mathematical role in the following, despite the presence of long range forces. For a homogenous interaction, we  abbreviate it with
\begin{equation}\label{W_0}
W_{nn}=W_0.
\end{equation}
With the many-body model above, we will consider controlled interaction quenches to illustrate our results in \sref{extendedBH}. These are readily realized with e.g.~Rydberg atom interactions due to the possibility of time-dependent external control. 

\subsection{Rydberg excitations in a Bose-gas}
\label{rydberg_preview}

As a second example, we will consider a Bose-gas where atoms are continuously coherently excited from a groundstate $|g \rangle$, in which interactions are neglected, to a Rydberg state $|e \rangle$ with strong long-range interactions between the atoms \cite{singer:blockade,tong:blockade,heidemann:strongblockade, johnson:rabiflop,tanner:manybodyionization, raitzsch:echo}. Initially all atoms are in the ground-state. With more atoms getting excited, the so-called dipole blockade develops \cite{singer:blockade,tong:blockade,heidemann:strongblockade}, where the simultaneous excitation of nearby atoms is strongly suppressed.

For this system we use a continuum formulation of the Hamiltonian \bref{Hamiltonian}. To this end, we define field operators via $\aop{n}=\hat{\Psi}(\mv{x}_{n})\sqrt{dV}$, $\aop{m}=\hat{\Psi}(\mv{y}_{m})\sqrt{dV}$, $W_{nm}=W(\mv{x}_n-\mv{y}_m)$, 
and $\omega_{nm}=-(\delta_{n+1,m} + \delta_{n-1,m} -2 \delta_{n,m})/(2m\: dV^2)$. Here, $\hat{\Psi}(\mv{x})$ is a field operator creating an atom at a location $\mv{x}$, $dV$ is the (physically small) volume corresponding to one discrete location $\mv{x}_n$ on the numerical lattice. We set $\hbar=m=1$ for simplicity, unless stated otherwise. 

We then extend our model to include two fields describing atoms in components $|g \rangle$ and $|e \rangle$. Only the latter shall experience the long-range interactions. 
Rydberg excitation and dynamics often happen much faster than atomic motion, which justifies the so-called ``frozen gas'' approximation where atomic motion is neglected. 
With the coherent coupling amplitude $\kappa$ between the components we then have:
\begin{eqnarray}
\hat{H}=\frac{\kappa}{2} \int d\mv{x}\left( \psidag{e} \psiop{g}+ \psidag{g} \psiop{e}\right) 
\CR
+ \frac{1}{2}  \int \!d\mv{x} \!\! \int \!d\mv{y} \: \psidag{e} \Psi^{\dagger}_{e}(\mv{y})W(\mv{x}-\mv{y})  \psiop{e} \Psi_{e}(\mv{y}).
\label{Hamiltonian_Ryd}
\end{eqnarray}
The crucial difference to the extended Bose-Hubbard model is that the number of atoms in the component experiencing the interactions is no longer conserved. We will assume the same interaction potential $W$ as in the previous model \bref{potential}.

For short times, \eref{Hamiltonian_Ryd} represents an adequate model (see e.g.~\cite{moebius:bobbels}). However corrections due to atomic motion can easily become relevant \cite{wuester:cradle,wuester:CI,leonhardt:switch,leonhardt:unconstrained,thaicharoen:trajectory_imaging,thaicharoen:dipole_trajectory_imaging,celistrino_teixeira:microwavespec_motion}, in which case the kinetic-energy term
\begin{eqnarray}
\hat{H}_{\rm kin}= -\int dx \left( \psidag{e} \frac{\nabla^2}{2 m }\psiop{e} + \psidag{g} \frac{\nabla^2}{2 m }\psiop{g} \right)
\label{Hamiltonian_kin}
\end{eqnarray}
must be reinstated to \eref{Hamiltonian_Ryd}, precluding the use of a spin model. This reduces the range of methods available, but does not constitute a problem for phase-space methods. 

With or without motion, we will later be interested in correlation function dynamics at the onset of a dipole-blockade: Assume all atoms are initially condensed in the groundstate. The coupling $\kappa$ then acts for a given time, transferring population into the excited state, which has interactions between atoms. The interaction energy of atoms in the excited state can become large enough to suppress transition to that state after an initial one. As a result, there is only one atom excited per ``blockade volume''. The Rydberg atom autocorrelation function $g^{(2)}$ has a pronounced dip for radii less than the blockade radius $r_{b}$ and the frequency of incomplete Rabi oscillations increases to $\kappa_{mb}=\sqrt{N_{b}}\kappa$, where $N_{b}$ is the number of ground state atoms within the blockade volume and $\kappa$ is the single atom Rabi frequency~\cite{lukin:quantuminfo}. 

\section{Gauge-P and positive-P descriptions}
\label{method}

If we consider $M$ single particle modes $\ket{m}$, populated with up to $N$ particles, the number of many-body basis states of our problem scales like $M^{N}$, an astronomical number for all but the smallest systems. The first-principles simulation of the Hamiltonian \bref{Hamiltonian} is thus a severe challenge. 
We now briefly review how the dynamics can be tackled with the gauge-P distribution, by stochastically sampling the many-body state, rather than representing it exactly.
 For an in-depth explanation we refer to the extensive literature on the Positive-P 
\cite{drummond:gardiner:posp,steel:wigner,book:qn,drummond:review:jmo,drummond99,craig:dissoc,deuar:collisions,Polkovnikov10,stab,Ng11,gardiner:boundaries,Carter87,deuar:jphysAI,deuar09,swistak16}
and Gauge-P methods 
\cite{plimak:diffusion,deuar:pra,Carusotto01a,Carusotto01b,deuar:job,deuar:jphysAII,deuar:jcpc,deuar:thesis,Aimi07b,deuar:canonical,deuar09b,Dowling07,Ng13}.

\subsection{Phase-space representation}
\label{rep}
First, we expand the system's density matrix $\hat{\rho}$ in an over-complete operator basis of coherent states:
\begin{eqnarray}
\hat{\rho}(t)=\int d^{2M}\balpha \int d^{2M}\bbeta \int d^{2}\Omega \:\: P_G(\balpha,\bbeta,\Omega,t) \left[ \Omega
\frac{| \balpha \rangle\langle \bbeta^{*}|}
{\langle \bbeta^{*}|\balpha\rangle} \right].
\label{GaugePexpansion}
\end{eqnarray}
The operator kernel in this expression is $\hat{\Lambda}=\Omega | \balpha \rangle\langle \bbeta^{*}|/\langle \bbeta^{*}|\balpha\rangle$, where we have made use of many-mode coherent states\footnote{The many-mode coherent states are a separable tensor product of coherent states $\otimes_{n=1}^M |\alpha_{n}\rangle$ with complex amplitude $\alpha_{n}$ for each of the single particle modes.} $|\balpha\rangle$ defined by $\aop{n}|\balpha\rangle=\alpha_{n}|\balpha\rangle$, with $\alpha_n\in {\mathbb C}$. The function $P_G$ is the ``Gauge-P'' distribution, which can be chosen positive and real for any density matrix $\hat{\rho}$, similarly to the positive-P distribution $P_+(\balpha,\bbeta,t)$, which is the special case that occurs when the global weight $\Omega$ is chosen to be constant.

For unitary dynamics, on which we will focus here, the density matrix obeys the von Neumann equation
\begin{eqnarray}
\frac{d }{d t}\hat{\rho}=-\frac{i}{\hbar}[\hat{H},\hat{\rho}].
\label{mastereqn}
\end{eqnarray}
Dissipative environments that would add, for example, Lindblad terms to \eref{mastereqn} and produce a master equation, can be straightforwardly included.

Due to the doubling of the phase space dimension for quantum systems seen above, it will be convenient to introduce the following notation: 
We collect all $2M+1$ stochastic fields into a vector $\underline{\bv{v}}=(\balpha, \bbeta, \Omega)$ whose components  $v_{\mu}$  are labeled by greek indices $\mu,\nu,\dots$, as opposed to the $M$ modes which are labeled by roman indices $n,m,j,\dots$.  Continuing this distinction for vectors in general, bold-face, underlined vectors will have $2M$ components\footnote{or $2M+1$ components if the weight $\Omega$ is included.}, without the underline they have $M$ components. 
Similarly a double underline is used to distinguish $2M\times2M$ matrices from the $M\times M$ matrices that they are usually composed of. Components of extended matrices have greek subscripts that run over $2M$ values.

For a sufficiently well bounded distribution $P_G(\balpha,\bbeta,\Omega)$, the master equation can be converted into a Fokker-Planck equation (FPE) of the generic form 
\begin{eqnarray}
\frac{\partial P_G}{\partial t}=
-\sum_{\mu} \pdiff{}{v_{\mu}}A_{\mu}P_G 
+\frac{1}{2}\sum_{\mu\nu}
\pdiff{}{v_{\mu}}\pdiff{}{v_{\nu}}
D_{\mu\nu}P_G
\label{generic_fpe}
\end{eqnarray}
by virtue of the operator correspondences~\cite{joe:superchem,deuar:thesis}:
\begin{subequations}
\begin{eqnarray}
\aop{n} \hat{\rho}\leftrightarrow \alpha_{n} P_G, 
\:\:\:\:
\adag{n}  \hat{\rho} \leftrightarrow \left(\beta_{n} - \frac{\partial}{\partial \alpha_{n}} \right)P_G,
\label{replacement_rules1}
\\
\hat{\rho} \aop{n}  \leftrightarrow \left(\alpha_{n} - \frac{\partial}{\partial \beta_{n}} \right)P_G,
\:\:\:\:
\hat{\rho}\adag{n}  \leftrightarrow \beta_{n} P_G,
\label{replacement_rules2}
\\
0 \leftrightarrow P_G +  \pdiff{}{\Omega}   \Omega P_G.
\label{replacement_rules3}
\end{eqnarray}
\end{subequations}

Crucially, \eref{generic_fpe} is equivalent to the set of coupled stochastic differential equations (SDEs)
\begin{eqnarray}
\frac{d v_{\mu}}{dt}=A_{\mu} + \sum_{\nu}B_{\mu\nu}\,\xi_{\nu}(t),
\label{generic_sde}
\end{eqnarray}
if the \emph{diffusion matrix} $\mvv{D}$ is positive semi-definite
\footnote{A positive semi-definite diffusion matrix has all eigenvalues non-negative, and such a form for $D$ can always be constructed here due to $\hat{\Lambda}$ being an analytic function of complex variables \cite{drummond:gardiner:posp}.}
~\cite{book:qn,book:stochmeth}. 
These SDEs have to be interpreted in the It{\^o}-form of stochastic calculus~\cite{book:stochmeth}.
The $\xi_{\nu}(t)$ are independent real noise fields, defined by a zero mean $\overline{\xi_{\nu}(t)}$ and correlations
$\overline{\xi_{\mu}(t)\xi_{\nu}(t')}=\delta_{\mu,\nu}\delta(t-t')$, where $\overline{f}$ denotes the stochastic average of $f$. They are usually implemented with independent Gaussian noises at each time step $\Delta t$ with a variance of $1/(\Delta t \:dV)$.
The ``noise matrix'' $\mvv{B}$ must satisfy $\mvv{D}=\mvv{B}\mvv{B}^{T}$. The matrix square root $\mvv{D}^{1/2}$ is a viable noise matrix in most cases of interest, provided that it is self-transposed ($\mvv{D}^{1/2}=[\mvv{D}^{1/2}]^T$).

All quantum expectation values (observables) can be found in principle by averaging
\footnote{Pathological cases can occur if the distribution $P_G$ does not decay sufficiently rapidly for large $|\alpha|$, $|\beta|$. This is discussed e.g. in \cite{gardiner:boundaries,deuar:pra,DeuarPhD-bt}.}
 over the set of independent stochastic trajectories. Expectation values of normal ordered operator products take the form~\cite{deuar:pra}
\begin{eqnarray}
\movel
\langle \hat{a}^{\dagger}_{n} \hat{a}^{\dagger}_{m} \hat{a}^{\dagger}_{l} \dots \hat{a}_{p} \hat{a}_{q} \hat{a}_{r} \dots \rangle=\frac{\overline{\Omega \beta_{n}\beta_{m}\beta_{l} \dots \alpha_{p} \alpha_{q} \alpha_{r} \dots  +   
\Omega^* \alpha_{n}^* \alpha_{m}^* \alpha_{l}^* \dots \beta_{p}^* \beta_{q}^* \beta_{r}^* \dots
   }}{\overline{\Omega + \Omega^*}}.
\label{observables}
\end{eqnarray}
The discussion above also applies to the Positive-P distribution, in which case we can set $\Omega\to1$ and simplify the calculation considerably \cite{deuar:jphysAI}. 

Upon reaching \eref{generic_sde}, we have reduced the generally intractable quantum-many body problem in an $N^M$ dimensional Hilbert-space to the seemingly much easier solution of a system of  $2M$ coupled complex SDEs. 
Whether the problem in this new form is tractable depends on the number of stochastic realizations required to obtain converged results for \eref{observables}. Therein lies the fundamental limitation of stochastic phase-space methods: After a certain simulated time $t_{\rm sim}$, the noise in the simulation can become nonlinearly amplified which prevents convergence of the averages. As a consequence, these stochastic methods are only useful for understanding properties of the system that manifest themselves before $t_{\rm sim}$.

Fortunately, the above derivation of the equations of motion contains several mathematical degrees of freedom, that do not affect the simulated physics.
These degrees of freedom are termed ``gauges''. Under favorable conditions, they can be exploited to limit the growth of noise and obtain simulations that give results for longer times $t_{\rm sim}$. The following options for gauges are available:
Firstly, we can use the replacement rule \bref{replacement_rules3} to add functions of our choice to the \emph{drift vector} $A_{\mu}$ for the evolution of the $\balpha$ and $\bbeta$. These functions are the \emph{drift gauges}. As a trade-off, their use induces a stochastic evolution of the global trajectory weight $\Omega$. Secondly, we can transform the noise matrix $\mvv{B}$ via $\mvv{B}'=\mvv{B}\mvv{O}$ with an orthogonal matrix $\mvv{O}$ such that $\mvv{O}\mvv{O}^T=I$. If $\mvv{B}$ had fulfilled the diffusion condition $\mvv{D}=\mvv{B}\mvv{B}^{T}$, then clearly so does $\mvv{B}'$. An orthogonal matrix $\mvv{O}$ can be written as $\mvv{O}=\exp{\mvv{G}}$, where $\mvv{G}$ is anti-symmetric ($\mvv{G}^{T}=-\mvv{G}$). It has been shown that the real part of $\mvv{G}$ merely induces inconsequential re-summations of the noise, whereas the imaginary part redistributes the noise between amplitude and phase components of the stochastic fields and can have strong influence on the available simulation times~\cite{deuar:pra,deuar:job,deuar:thesis,plimak:diffusion}. The matrices $\mvv{G}$ or $\mvv{O}$ are called \emph{diffusion gauges}.

Different gauges yield different stochastic equations of motion, whose solutions have different convergence properties. Nonetheless, the evolution still corresponds to one and the same physical density matrix, and averages such as \bref{observables} with physical meaning remain unique. Thus gauges affect the mathematical but not the physical properties of our problem, just as their namesake in electro-dynamics.

\subsection{Direct form of the equations of motion}
\label{gaugep_eoms}
Now we apply this formalism to the lattice system described by Hamiltonian \bref{Hamiltonian}. 
In the positive-P representation (without gauges), the diffusion matrix is block-diagonal with separate blocks for the $\alpha$ and $\beta$ variables~\cite{book:qn, deuar:thesis}., i.e. 
\begin{equation}
\mvv{D} =\left( \begin{array}{cc}
D^{(\alpha)} & 0 \\
0 &D^{(\beta)}
\end{array} \right), \qquad
D^{(\alpha)}_{nm} = -i W_{nm}\alpha_n\alpha_m,\qquad 
D^{(\beta)}_{nm} = i W_{nm}\beta_n\beta_m.\qquad
\label{Dpp}
\end{equation}
Realistic two-body interactions are symmetric, $W_{nm}=W_{mn}$, which leads to symmetric roots $\sqrt{W}=W^{1/2}=[\sqrt{W}]^T$. Then, the simplest decomposition of the noise matrix $\mvv{D}$ gives the following equations:
\begin{subequations}\label{ppequations}
\begin{eqnarray}
\frac{d\alpha_n}{dt}&=-&i\sum_m \left[\omega_{nm}\alpha_m + \alpha_n W_{nm}\alpha_m\beta_m\right] - i\sqrt{i}\,\alpha_n\sum_j [\sqrt{W}]_{nj} \xi^{(1)}_j(t),\mbox{\hspace{0.6cm}}\\
\frac{d\beta_n}{dt}&=&i\sum_m \left[\omega_{nm}\beta_m + \beta_n W_{nm}\alpha_m\beta_m\right] + \sqrt{i}\,\beta_n\sum_j [\sqrt{W}]_{nj} \xi^{(2)}_j(t).
\end{eqnarray}
\end{subequations}
The noises $\xi^{(1)}_j$ and $\xi^{(2)}_j$ are independent, and $[\sqrt{W}]_{nj}$ are components of the matrix square root $\sqrt{W}$ of the interaction potential. 

For the gauge-P representation, with stochastic gauge freedoms included, it will be convenient to use the matrix notation described at the beginning of section~\ref{rep} to conscisely deal with the $2\times2$ block nature of the problem. Vectors are assumed to be column vectors unless explicitly transposed, and we use the notation $\mbox{diag}[\bv{v}]$ do assemble a square diagonal matrix with vector $\bv{v}$ on the diagonal. 
Following \cite{joel:gauss1,joel:fermbos:prl}, we introduce the following:
\begin{eqnarray}
\bvv{\gamma}=\left( 
\begin{array}{c}
\bv{\alpha}
\\
\bv{\beta}
\end{array} 
\right),\:\:\:
\mvv{\omega} =\left( 
\begin{array}{cc}
-\omega & 0 \\
0 &\omega
\end{array} 
\right),\:\:\:
\mvv{W}=\left( 
\begin{array}{cc}
-W & 0 \\
0 &W
\end{array} 
\right),\:\:\:
\nonumber\\
\bvv{n}=\left( 
\begin{array}{c}
\bv{n}
\\
\bv{n}
\end{array} 
\right)
=
\left( 
\begin{array}{c}
(\dots,\alpha_{j}\beta_{j},\dots)^{T}
\\
(\dots,\alpha_{j}\beta_{j},\dots)^{T}
\end{array} 
\right),\:\:\:
\bvv{A} = i \mvv{\omega}\bvv{\gamma} + i \mvv{\Gamma}\mvv{W}\bvv{n},
\label{unified_eoms_posp}
\\
\bvv{\xi}=\left( 
\begin{array}{c}
\bv{\xi}^{(1)}
\\
\bv{\xi}^{(2)}
\end{array} 
\right),\:\:\:
\mvv{\Gamma}=\mbox{diag}(\bvv{\gamma}),
\:\:\:
\mvv{B}=\sqrt{i} \mvv{\Gamma} \mvv{S},\:\:\:
\mvv{S}=\left( 
\begin{array}{cc}
-i\sqrt{W} & 0 \\
0 & \sqrt{W}
\end{array} 
\right)\!.\nonumber
\end{eqnarray}
A particularly useful quantity that will recur is the (complex) mode occupation
\begin{equation}
\label{ndef}
n_{m} = n'_m+in''_m= \alpha_m\beta_m.
\end{equation}
Using these definitions, we can write the Gauge-P representation equations of motion
\begin{subequations}
\begin{eqnarray}
\pdiff{}{t}\bvv{\gamma}=\bvv{A} + \mvv{B'}(\bvv{\xi} -\bvv{g}),\:\:\:\:\:\:
\pdiff{}{t}\Omega= \Omega\, \bvv{g}\cdot \bvv{\xi},\label{gaugev}
\\
\mvv{B'}= \mvv{B} \mvv {O}=\sqrt{i} \mvv{\Gamma} \mvv{S} \mvv{O},\:\:\:\:\:\:
\bvv{g}=\left( 
\begin{array}{c}
\bv{g}^{(1)}
\\
\bv{g}^{(2)}
\end{array} 
\right),
\end{eqnarray}
\label{unified_eoms_gaugep}
\end{subequations}
%
as per the formalism outlined in \cite{plimak:diffusion,deuar:pra,deuar:job,deuar:jphysAII, deuar:jcpc,deuar:thesis,deuar:canonical,deuar:collisions}. 

The \emph{drift-gauge} $\bvv{g}$ is a yet unspecified function of the variables $\bvv{\gamma}$. We also have already inserted an arbitrary orthogonal matrix $\mvv{O}=\exp[\mvv{G}]$  with $\mvv{O} \mvv{O}^{T} =\mvv{\id}$, the \emph{diffusion gauge}, into the diffusion term.

\section{Optimised gauges}
\label{optimalgauges}

For practical applications of the available stochastic gauges, we introduce more specific forms and then optimize them, similarly to how it was done for local interactions in \cite{deuar:jphysAI,deuar:jphysAII}. 

\subsection{Intuitive drift gauge form}
\label{gaugep_driftgauges}

Here, we introduce a specific form of $\bv{g}$ and describe its effect on the equations of motion. The objective of $\bv{g}$ is to control the tendency of the interaction term to create so-called moving instabilities and boundary term errors in the distribution $P_G$~\cite{deuar:pra,deuar:job,deuar:thesis}. Let
\begin{eqnarray}
g_{\mu}=\sqrt{i}\sum_{\nu \sigma} O_{\nu\mu}S_{\sigma\nu} f_{\sigma},
\end{eqnarray}
where $\bv{f}$ is a yet unspecified vector that becomes the gauge. In matrix notation this reads $\bvv{g}=\sqrt{i}\,\mvv{O}^{T}\mvv{S}^{T} \bvv{f}$. Using the properties of $\mvv{O}$ and $\mvv{S}$,such as $\mvv{S}\mvv{S}^T=\mvv{S}\mvv{S}=\mvv{W}$, one can see that the new equations of motion are
\begin{subequations}
\begin{eqnarray}
\pdiff{}{t}\gamma_\mu=i \sum_{\nu} \omega_{\mu \nu} \gamma_{\nu} + i\sum_{\nu}\gamma_{\mu}W_{ \mu \nu}(n_{\nu} -f_{\nu}) + \sqrt{i}\sum_{\nu\sigma} \gamma_{\mu}S_{\mu \nu}O_{\nu\sigma}\xi_{\sigma},
\label{final_stoch_eqalphabeta}
\\
\pdiff{}{t}\Omega= \sqrt{i}\Omega \sum_{\nu\sigma\lambda} f_{\lambda}O_{\sigma\nu}S_{\lambda\sigma}\xi_{\nu}=\sqrt{i}\,\Omega\,\bvv{f}^T \mvv{S} \mvv{O} \bvv{\xi}. 
\label{final_stoch_eqOmega}
\end{eqnarray}
\label{final_stoch_eqGaugeP}
\end{subequations}
Using the following notation for the real and imaginary part of complex numbers: $z=z' +iz''$, we can now pick for instance $f_{\lambda}=i n_{\lambda}''$ to remove the imaginary part of the stochastic density $\bvv{n}$ from the non-linear interaction term of \eref{unified_eoms_gaugep}, or $f_{\lambda}=n_{\lambda} - |n_{\lambda}|$ to replace $\bvv{n}$ by its modulus. Both choices reduce the tendency of the nonlinear term in \eref{unified_eoms_gaugep} to result in exponentially diverging trajectories \cite{deuar:pra}. 

In the following sections, we will make the choice 
\begin{equation}\label{driftg}
\bvv{f} = i \bvv{n}''
\end{equation}
for the drift gauge. This is a form that was found to be particularly useful for the dynamics of the contact-interacting gas in previous work, giving the best extensions of simulation time \cite{deuar:jphysAII}.

Most of the efficient numerical time-propagation schemes for stochastic differential equations require the use of the Stratonovich form. The necessary corrections to convert \eref{final_stoch_eqGaugeP} 
from the It\^o to the Stratonovich form are described in \aref{stratocorrapp}.

\subsection{Ensemble spread and characteristic variance}
\label{noiseevol_GaugeP}

To use the diffusion gauge beneficially, we have to know how a given choice for the matrix $\mvv{O}$ affects the noise induced spread of trajectories in the ensemble as time progresses. 
The trajectory spread determines the simulation time available. 
A characteristic variance ${\cal V}_1=\frac{1}{2}(\mbox{var}[\log|\Omega\alpha|] + \mbox{var}[\log|\Omega\beta|])$
was used for the single mode case in the past to obtain quite accurate estimates of the simulation time $t_{\rm sim}$ \cite{deuar:jphysAII}. 
The quantities $\Omega\gamma$ appearing  in ${\cal V}_1$, are the estimators that appear in \eref{observables} for expectation values of the one-point correlation function. They are fundamental to most quantities that one might want to calculate. The logarithm is used in anticipation of an exponential increase in noise variances throughout the simulation, as is observed in practice. 
The first task here, and a keystone of subsequent analysis,
is to generalize this quantity for our purposes. 

A many mode generalization of ${\cal V}_1$ is
\begin{eqnarray}
{\cal V}=\frac{1}{2M}\sum_{\mu}^{2M}\mbox{var}\left[\log|\Omega \gamma_{\mu}|\right].
\label{charact_variance}
\end{eqnarray}
The stochastic uncertainty of fields in all modes contributes to ${\cal V}$, hence this simple scalar quantity can signal when the simulation becomes too noisy to extract useful information via \eref{observables}. 
Even if just the variance of $\gamma_{\mu}$ for a specific mode $\mu$ becomes too large, this will soon contaminate all the remaining modes through the linear coupling terms $\propto \omega$ in the equations of motion. Note, that ${\cal V}$ does not have any physical interpretation, and indeed must not, since it should be gauge-dependent.

For the ``standard'' drift gauge \bref{driftg}, the approximate time evolution of \eref{charact_variance} is calculated from \eref{final_stoch_eqGaugeP},  in \aref{charvarapp}. 
The calculation neglects the kinetic $\omega_{\mu\nu}$ part of the drift, as was done for the two-mode case in \cite{deuar:jphysAII}, expecting the predominant contribution to the increase of ${\cal V}$ to come from the noise terms and non-linearities. The remaining terms conserve the expectation value of the local density $\langle n_{m}\rangle$, so that ${\cal V}(t)$ depends only on their initial values $n_{m}(0)$.
The result is
\begin{subequations}
\label{variance_final}
\begin{eqnarray}
\movel
{\cal V}(t)= {\cal V}(0) + \frac{1}{2M}\Bigg\{
\frac{t}{2}\mbox{Tr}\left[\mvv{S} \mvv{O} \mvv{O}^{\dagger} \mvv{S}^{\dagger} \right]  
+t\, \mbox{Tr}\left[
 {\rm Im}\left\{ \left[\mvv{S} \mvv{O}\mvv{O}^{\dagger} \mvv{S}^{\dagger}\right]\right\} \mvv{N}''\right]
\CR
+M\mbox{Tr} \bigg[{\rm Re}\left\{\mvv{S} \mvv{O}\mvv{O}^{\dagger} \mvv{S}^{\dagger} \right\} \mvv{Q}
 \bigg]\Bigg\},
\label{variance_final_expr}
\\
\movel
Q_{\mu\nu}=\frac{1}{2}{\rm Re}\left[\overline{n_{\mu}(0)n^{*}_{\nu}(0)}\Big(\exp\left[t\,P_{\mu\nu}\right] -1 -t\,P_{\mu\nu}\Big)/P_{\mu\nu}\right] + t\,[\overline{n''_{\mu}(0)n''_{\nu}(0)}]
\label{Qresult}
\\
\movel
P_{\mu\nu} = \left[\mvv{F} \mvv{S} \mvv{O}\mvv{O}^{\dagger}\mvv{S}^{\dagger}\mvv{F}\right]_{\mu\nu}.
\label{Pdef}
\end{eqnarray}
\end{subequations}
This is central to most of the rest of the paper. 
A useful $2M\times 2M$ matrix is
\begin{equation}
\mvv{F}=\left( 
\begin{array}{cc}
\id & \id \\
\id & \id
\end{array} 
\right),
\label{Fdef}
\end{equation}
and $[\cdots]_{\mu\nu}$ denotes the component $\mu\nu$ of the matrix product within square brackets. 
The auxilliary diagonal matrix $\mvv{N}$ is 
\begin{equation}
\mvv{N} = \mvv{N}'+i\mvv{N}'' = \mbox{diag}[\overline{\bvv{n}(0)}].
\end{equation}
and has the initial density on the diagonal.

The first term of \eref{variance_final_expr} is the contribution from noise placed directly into the amplitudes $\bvv{\gamma}$.  The third term, involving $\mvv{Q}$, arises from noise accumulated in the weight $\Omega$.  The remaining term, involving $\mvv{N}''$, comes from an accumulation of noise in $\Omega$ due to any initial fluctuations in $\bv{n}''(0)$ and is less important.. For single-mode repulsive contact interactions we have $W_{nm}=(g/dV)\delta_{nm}>0$, and thus \eref{variance_final_expr} reduces to the known results \cite{deuar:jphysAII}.

The variance \eref{variance_final} forms the essential basis for most other results of this article along with \eref{variance_finalP}, and may additionally contain useful information for future work on stochastic gauges. The calculation is involved but indispensable to (i) determine simulation time limitations that help to assess whether the methods here are applicable to a given physical problem, (ii) assess whether gauges are expected to outperform the direct application of the (ungauged) Positive-P method, (iii) operate diffusion gauges in analogy to the single mode case and (iv) possibly enable future work to fully harness the large number of gauge degrees of freedom available in the long-range interacting many-mode case.

\subsection{Global diffusion gauge}
\label{globalDG}

To make use of \eref{variance_final} to choose a good gauge, we have to specify the form of the diffusion gauge matrix $\mvv{O}$. In the single mode case~\cite{plimak:diffusion,deuar:jphysAII} available simulation times for the Gauge-P method can be substantially improved using $\mvv{O}=-\cosh(a) \sigma_{3} -i \sinh(a)\sigma_{1}$, where $a$ is an adjustable real \emph{gauge parameter} and $\sigma_{i}$ are the standard Pauli matrices. The underlying mechanism is a shift of the numerical noise from the amplitude of $n=\alpha \beta$ to its phase, which less directly 
leads to deleterious noise amplification. 
The simplest generalisation to many modes is to use one global value of the parameter  $a$, giving the \emph{global diffusion gauge}
\begin{eqnarray}
\mvv{O}=\left( 
\begin{array}{cc}
\cosh a\id & -i\sinh a\id \\
i\sinh a\id &  \:\:\cosh a\id
\end{array} 
\right)=\exp(g), \:\:\:\:\:\:\
g=\left( 
\begin{array}{cc}
0& -i  a\id \\
i a\id &0
\end{array} 
\right).
\label{globalO}
\end{eqnarray}
For a nonuniform system (in a trap, etc.), the obvious adaptation is to have a vector $\bv{a}$ of gauge parameters with spatially varying values $a_n=a(\mv{x}_n)$. 
Such a gauge could most easily be implemented piecewise if the length scale on which the density varies is larger than the characteristic lengthscale of the potential $W$. Then each region would correspond to a block like \eref{globalO}, optimised using densities inside the block. A local gauge with arbitrarily varying $a(\mv{x})$ seriously complicates analysis, because there is an interplay between the lengthscales of the density variation and of the interaction. We will comment on several such  more involved choices for the gauge in \sref{nonlocal_diff}. 

Returning to the global choice \eref{globalO}, the quantities appearing in \eref{variance_final} simplify and are
\begin{eqnarray}
\movel
\mvv{O}\mvv{O}^{\dagger} = 
\left(\begin{array}{cc}
\:\id\cosh2a & -i\id\sinh2a \\
i\id\sinh2a &  \:\:\id\cosh2a
\end{array}\right),
\qquad
\mvv{P} = 2e^{-2a}
\left(\begin{array}{cc}
U&U\\
U&U
\end{array}\right).
\end{eqnarray}

\subsubsection{Noise minimisation and gauge choice}
\label{gaugechoice}

Assuming the form \eref{globalO}, an optimal parameter $a$ can be found by minimizing~${\cal V}(t_{\rm opt})$, where $t_{\rm opt}$ is the chosen target time, typically the longest time for which simulation results are required \cite{deuar:jphysAII}. 
The analytical minimization of \eref{variance_final} is only tractable with several simplifications. 
We proceed in similar fashion as was done in \cite{deuar:jphysAII} for a single-mode case.
We expand the variance \eref{variance_final} to second order in $t$:
\begin{eqnarray}
\movel
{\cal V}(t)={\cal V}(0) + \frac{t}{2}U_0\cosh{(2a)}+t e^{-2a} I_{1} +\frac{t^{2}}{2}e^{-4a}I_{2},
\label{variance_localO5}
\\
\movel
I_{1}\equiv \sum_{kk'}U_{kk'} \overline{n''_{k}(0)n''_{k'}(0)}  \approx \int d\mv{x}\, d\mv{y}\, \overline{\rho_{0}''(\mv{x})\rho_{0}''(\mv{y})}\,U(\mv{x}-\mv{y}),
\CR
\movel
I_{2}\equiv \sum_{kk'} U_{kk'}^2 \mbox{Re}[\overline{n_{k}(0)n^*_{k'}(0)}]  \approx \int d\mv{x}\, d\mv{y}\, \overline{\rho_{0}(\mv{x})\rho_{0}^*(\mv{y})}\,U(\mv{x}-\mv{y})^2,
\nonumber
\end{eqnarray}
where, for the continuum version of the integrals, we define the complex density $\rho_{0}(\mv{x})=\rho'_{0}(\mv{x})+i \rho''_{0}(\mv{x})=n(\mv{x}_n)/dV$. Here, we have also introduced a semi-positive definite ``rectified interaction'' matrix 
\begin{equation}
\label{Udef}
U = \sqrt{W}\sqrt{W}^{\dagger}
\end{equation}
that plays an important role and will often reappear later.
In general, its diagonal elements $U_0=U_{nn}$ are not equal to those of $W$ which are $W_0$, but they can be related, and are always positive real
\footnote{Writing the potential $W$ as an eigenvalue decomposition with orthogonal normalized eigenvectors $|j\rangle$ such that $\langle j|k\rangle=\delta_{jk}$ 
we have that  $W=\sum_j \lambda_j |j\rangle\langle j|$ and hence $U=\sum_j |\lambda_j|\, |j\rangle\langle j|$.
For cases when the potential $W$ has all eigenvalues positive/negative, one thus finds that $U_0=|W_0|$. Otherwise, $U_0>|W_0|$, and the relationship is more complicated.  
}.
The potential \eref{potential} used for the examples in section \ref{MBsim}, \emph{usually} has all eigenvalues negative though not always, particularly if periodic boundary conditions occur.

It is instructive to consider the limit of contact interactions, defining their strength $g$ via $U_0\to|g|/dV$.  The integrals $I_j$ in \bref{variance_localO5} then become
\begin{equation}
\movel
I_1\to MU_0\overline{(n''(0))^2}\sim|g|(\rho_0'')^2\,V,
\qquad\mbox{and}\qquad 
I_2\to MU_0^2\overline{|n(0)|^2}\sim \left[\frac{g^2|\rho_0|^2}{dV}\right]\,V.\quad
\label{Iest}
\end{equation}
We see that the $I_j$ are extensive quantities, growing with system size, whether measured by mode count $M$ or volume $V$. 

We now wish to minimize ${\cal V}(t_{\rm opt})$
at a given time $ t_{\rm opt}$. Following \cite{deuar:jphysAII}, we obtain analytically tractable results in two limits:\\
{\it Weight dominated,}
\begin{eqnarray}
\movel\label{aopt1}
a_{\rm opt} \approx \frac{1}{6}\log\left(\frac{4t_{\rm opt}I_2}{U_0}\right),
\end{eqnarray}
when the $I_1$ term is small, i.e. $n_0$ and $t_{\rm opt}$ are large
and {\it direct noise dominated,}
\begin{eqnarray}
\movel\label{aopt2}
a_{\rm opt} \approx \frac{1}{4}\log\left(1+\frac{4I_1}{U_0}\right),
\end{eqnarray}
when the $I_2$ term is small, i.e. $n_0$ or $t_{\rm opt}$ is small.
Then we follow the same interpolation procedure that was successful in ~\cite{deuar:jphysAII} to give an overall best estimate
\begin{eqnarray}
a_{\rm approx}=\frac{1}{6}\log{\left(\frac{4 I_{2} t_{\rm opt}}{U_0} +\left[1 + \frac{4 I_{1}}{U_0} \right]^{3/2} \right)}.
\label{adaptive_local_dg}
\end{eqnarray}

\subsubsection{Adaptive diffusion gauge}
\label{adaptivegauge}
Single mode work has shown that the performance can sometimes be improved further, by adapting the gauge parameter $a$ in a time-dependent manner during the simulation, leading to an $a=a(t)$. 
In that case, one determines $a(t)$ at each timestep according to \eref{adaptive_local_dg} using $t_{\rm opt}=t_{\rm fin}-t$, where $t_{\rm fin}$ is the desired final time of the simulation and the values $n_{0}(\mv{x})$ appearing in \eref{adaptive_local_dg} are changed to the time-dependent $n(\mv{x},t)$. We then obtain
\begin{eqnarray}
a_{\rm adaptive}=\frac{1}{6}\log{\left(\frac{4 I_{2}[\bv{n}(t)](t_{\rm fin}-t)}{U_0} +\left[1 + \frac{4 I_{1}[\bv{n}(t)]}{U_0} \right]^{3/2} \right)}.
\label{adaptive_explicit}
\end{eqnarray}
The Stratonovich correction \eref{stratocorr} becomes more involved when $a$ depends on the phase-space variables $\gamma_{\mu}$, details of the derivation are given in \aref{adaptivestratocorrapp}. 

\section{Available simulation times}
\label{tsimsection}

Before attempting to simulate physics with the methods presented here, the crucial question is whether one expects the physical effects of interest to take place prior to materialize before the time $\sub{t}{sim}$ when the noise level overwhelms the simulation. As in earlier work \cite{deuar:jphysAII}, we define this moment via ${\cal V}(\sub{t}{sim})=10$. Based on the optimized gauges from the previous section, we will determine $\sub{t}{sim}$ for the Gauge-P method in this section.

Gauges do not always offer an advantage over the technically much simpler plain Positive-P method. In some other cases the use of diffusion gauges without drift gauge is preferable. To allow an a-priori assessment, we will also derive $\sub{t}{sim}$ for those methods. This requires a re-derivation of the characteristic variance for those cases, similar to \sref{noiseevol_GaugeP}, which is presented in the following.

\subsection{Positive-P and diffusion gauged characteristic variance}
\label{noiseevol_PosP}

Let us consider the case without drift gauges ($\bvv{g}=\bvv{f}=0$), which includes the plain positive-P representation as a special case ($\mvv{O}=\mvv{I}$), and will also be useful in \sref{diffusionG}. 
The approximate time evolution of $\cal V$ is calculated similarly to \eref{variance_final} from \eref{unified_eoms_gaugep}, taking $\Omega=1$ and $\bvv{f}=0$. Details can be found in \aref{charvarappP}. 
We obtain
\begin{subequations}
\label{variance_finalP}
\begin{eqnarray}
\movel
{\cal V}^{(P)}(t)= {\cal V}(0) + \frac{1}{2M}\Bigg\{ 
\frac{t}{2}\mbox{Tr}\left[\mvv{S} \mvv{O} \mvv{O}^{\dagger} \mvv{S}^{\dagger} \right]  
+ {\rm Tr}\left[\mvv{W} \widetilde{\mvv{Q}} \mvv{W}\right]
+t^2\:{\rm Tr}\left[\mvv{W} \mvv{C}^{(0)}\mvv{W}\right]
\CR
-\frac{t^2}{2}{\rm Tr}\left[ 
\mvv{W}\mvv{N}'\mvv{F}\mvv{W} - {\rm Im}\left(\mvv{S}\mvv{O}\mvv{O}^{\dagger}\mvv{S}^{\dagger}\mvv{N}^*\right) \mvv{F}\mvv{W} 
\right]
-2t\:{\rm Tr}\left[\mvv{W} \widetilde{\mvv{C}}^{(0)}\right]
\Bigg\}
\label{variance_finalP_expr}
\\
\movel
\widetilde{Q}_{\mu\nu} = 
{\rm Re}\bigg[\overline{n_{\mu}(0)n^{*}_{\nu}(0)}
\left(\exp\left[t\,P_{\mu\nu}\right] -1 -t\,P_{\mu\nu} -\frac{t^2\,P_{\mu\nu}^2}{2}\right)/P_{\mu\nu}^2\bigg].
\label{Qtilde}
\end{eqnarray}
\end{subequations}
In the above very general case, two initial state covariances appear,
\begin{eqnarray}
\label{C0}
\movel
C^{(0)}_{\mu\nu} = \mbox{covar}\left[ n''_{\mu}(0),\:\:n''_{\nu}(0)\,\right],
\quad\mbox{and}\quad
\widetilde{C}^{(0)}_{\mu\nu} = \mbox{covar}\left[ n''_{\mu}(0),\:\:\log|\gamma_{\nu}(0)|\,\right],
\end{eqnarray}
which are, however, zero in most cases.

The first term of \eref{variance_finalP_expr} is again the direct noise contribution placed into the amplitudes $\bvv{\gamma}$.  The next term involving $\widetilde{\mvv{Q}}$ is the main noise amplification, due to the nonlinearity working on the direct noise. The term involving $\mvv{N}'$ is a cross correlation between these, while the remaining terms are due to amplification of any initial fluctuations in $\bv{n}''(0)$. 

For the case of plain positive-P with $\mvv{O}=\mvv{1}$, \eref{variance_finalP} reduces to the following:
\begin{eqnarray}
\movel
{\cal V}^{(P)}(t)={\cal V}(0) + \frac{1}{2M}\Bigg\{
t\,\mbox{Tr}\left[U\right]  -t^2\,\mbox{Tr}\left[W N' W\right] + 2t^2\,{\rm Tr}\left[WC^{(0)}W\right] 
-2t\,{\rm Tr}\left[W\widetilde{C}^{(0)}\right]
\CR
+ \sum_{mkk'} W_{mk} \mbox{Re}\left[\frac{\overline{n_k(0)n^*_{k'}(0)}}{2}\left[ e^{2U_{kk'}t}-1-2U_{kk'}t -2(U_{kk'} t)^2\right]\,\right] W_{k'm}/U_{kk'}^2
\Bigg\}.\qquad
\label{charact_variance_PosP_oft}
\end{eqnarray}
We have used $M\times M$ matrices here. 

The initial covariances in \bref{charact_variance_PosP_oft} are
\begin{eqnarray}
\label{C0pp}
\movel
C^{(0)}_{kk'} = \mbox{covar}\left[ n''_{k}(0),\:\:n''_{k'}(0)\,\right],
\quad\mbox{and}\quad
\widetilde{C}^{(0)}_{kk'} = \mbox{covar}\left[ n''_{k}(0),\:\:\left(\log\left|\frac{\beta_{k'}(0)}{\alpha_{k'}(0)}\right|\right)\,\right].
\end{eqnarray}
The known single mode results ~\cite{deuar:jphysAI,deuar:jphysAII} for repulsive interaction are recovered from \eref{charact_variance_PosP_oft} with the replacement $M=1$, $W=(g/dV)\delta_{nm} > 0$, and var$[n''(0)]=0$.\,\footnote{
The calculations in \cite{deuar:jphysAI,deuar:jphysAII} had assumed a deterministic initial state such that $\overline{n(0)}=n(0)$, and thus did not contain the covariance contributions found here.}

\subsection{Useful simulation time (positive-P)}
\label{tsims_PosP}

As has been justified in detail in~\cite{deuar:jphysAI}, a characteristic variance of ${\cal V}\approx10$ sets the ultimate limit of usefulness in practice. It gives the variance of the \emph{logarithm} of a typical observable estimator, and exponentials of Gaussian random variables acquire intractably long distribution tails once this variance reaches a value of ${\cal O}(10)$. The observable estimators are largely of such a form.

Analytical results are only easily obtainable with several simplifications. We proceed in a similar fashion as was done in \cite{deuar:jphysAI} for a single-mode case.
First, we assume deterministic initial conditions so that the initial covariances $C^{(0)}$ and $\widetilde{C}^{(0)}$ can be neglected.
Then, we expand the variance \eref{charact_variance_PosP_oft} to third order in $t$. Unlike for \bref{variance_localO5}, here this is the first order that includes the important noise amplification contribution:
\begin{eqnarray}
\movel
{\cal V}^{(P)}(t)- {\cal V}(0)= + \frac{t}{2}U_0-\frac{t^2}{2}I^{(P)}_{1} +\frac{t^{3}}{3}I^{(P)}_{2}.
\label{charact_variance_PosP_oft_expanded}
\\
\movel
I^{(P)}_{1}\equiv \frac{1}{M}\sum_{kk'}W_{kk'}^2 \overline{n'_{k}}  
\approx \frac{1}{V}\int d\mv{x}\, d\mv{y}\, \overline{\rho_{0}'(\mv{x})}\,W(\mv{x}-\mv{y}),
\CR
\movel
I^{(P)}_{2}\equiv \frac{1}{M}\sum_{kk'k''} U_{kk'} W_{kk''}W_{k'k''} \mbox{Re}[\overline{n_{k}(0)n^*_{k'}(0)}]
\CR
  \approx \frac{1}{V}\int d\mv{x}\, d\mv{y}\, d\mv{z}\, \overline{\rho_{0}(\mv{x})\rho_{0}^*(\mv{y})}\,U(\mv{x}-\mv{y})W(\mv{x}-\mv{z})W(\mv{y}-\mv{z}).\nonumber
\end{eqnarray}
The rectified interaction $U$ and its diagonal value $U_0=U_{jj}\ge0$ appear again. 
 
To get an idea of the order of magnitude of these quantities, note that in the limit of contact interactions (as before, strength $g$, $U_0\to|g|/dV$, and density $\rho_0$), we have 
\begin{equation}
\movel
I^{(P)}_1\to W_0^2\overline{(n'(0))}\sim \frac{g^2\rho_0'}{dV},
\qquad\mbox{and}\qquad 
I^{(P)}_2\to U_0W_0^2\overline{|n(0)|^2}\sim \frac{|g|^3|\rho_0|^2}{dV}.
\label{IestP}
\end{equation}

The integrals $I_j$  are thus intensive quantities, independent of system volume, unlike those encountered in the corresponding Gauge-P derivation \bref{variance_localO5}. However they are dependent on the numerical lattice spacing $dV$, growing as lattice spacing drops. This is a manifestation of the known feature that simulation time improves as the lattice becomes more coarse.  

Consider now the useful case of coherent state (non-entangled) initial conditions, with $n''(0)=0$ and no starting variance (${\cal V}(0)=0$). We seek to determine when ${\cal V}^{(P)}=10$ is reached.   

The most important regime is when the last term in \eref{charact_variance_PosP_oft_expanded}, that comes from noise amplification, dominates the variance. By using the estimates \eref{IestP} to make analogy with the contact-interacting case we can get some bearings. This regime is seen to occur when
\begin{equation}
t|g|\rho_0 \gtrsim {\cal O}(1), \qquad\mbox{i.e.}\qquad t\gtrsim\frac{1}{|\mu|}
\end{equation}
for chemical potentials $\mu\approx g\rho$ that are typical for the superfluid regime of ultracold gases.
So, requiring ${\cal V}^{(P)}=10$ in this regime, one readily obtains 
\begin{eqnarray}
\movel
t_{\rm sim}\approx \frac{3}{\left[I^{(P)}_2\right]^{1/3}} \approx 3V^{1/3}\,\left[\int d\mv{x}\, d\mv{y}\, d\mv{z}\, \overline{\rho_{0}(\mv{x})\rho_{0}^*(\mv{y})}\,U(\mv{x}-\mv{y})W(\mv{x}-\mv{z})W(\mv{y}-\mv{z})\right]^{-1/3}
\label{tsimP}
\end{eqnarray}
This is a remarkably simple expression. It is roughly independent of the system size, and reduces to the known single-mode result 
$t_{\sim}\approx 2.5\,dV^{1/3} /[ g \rho_0^{2/3} ]$
in that limit \cite{deuar:jphysAI}. 

The opposite regime of short times or small chemical potential has evolution still dominated by the direct noise. Then the first term in \eref{charact_variance_PosP_oft_expanded} dominates and we have
\begin{eqnarray}
t_{\rm sim}\approx \frac{20}{U_0}.
\label{tsimP1}
\end{eqnarray}
The situation with dominant $I^{(P)}_1$ term does not cover a significant range of parameters.

\subsection{Useful simulation time (gauge-P)}
\label{tsims}

With an optimal gauge chosen as discussed in \sref{optimalgauges}, we can proceed to further estimate the useful simulation time $t_{\rm sim}$ for the gauge-P representation. We follow the same approach as in \sref{tsims_PosP} and \cite{deuar:jphysAII}. We take again coherent state initial conditions ($n''(0)=0$), and ${\cal V}(0)=0$. Then, $I_1$ from \eref{variance_localO5} vanishes, and we substitute the optimal gauges in the two regimes \bref{aopt1} and \bref{aopt2} into \eref{variance_localO5} and impose the condition ${\cal V}(t_{\rm sim})=10$. 

In the weight-dominated regime,  which applies for large systems or not-too-small densities, one obtains the important result that 
\begin{eqnarray}
t_{\rm sim}\approx \frac{8}{\sqrt{U_0 \sqrt{I_{2}}}},
\label{tsim3}
\end{eqnarray}
(assuming also $\cosh 2a\approx e^{2a}/2$). 
This reduces to  the results for the single-mode contact interacting case. In particular, for $M$ contact-interacting modes, 
\begin{eqnarray}
t_{\rm sim}\sim \frac{8\sqrt{dV}}{g\sqrt{\rho_0}}\times \frac{1}{M^{1/4}}.
\label{tsim3c}
\end{eqnarray} 
In the direct-noise dominated regime that is relevant for short times or small occupations,
\begin{eqnarray}
t_{\rm sim}\approx\frac{20}{U_0},
\label{tsim1}
\end{eqnarray}
which roughly applies when it is shorter than \eref{tsim3}, i.e. for $I_2\lesssim U_0^2/40$. Notably, it is the same as \eref{tsimP1} for positive-P. This criterion is usually only met for very small occupations $n_0$.

\section{Example simulations}
\label{MBsim}

We now proceed to demonstrate the utility of these methods with some exemplary applications.
The physical models to which we will apply the method have been presented in \sref{manybody}. For both examples, we consider for simplicity a one-dimensional (1D) system with periodic boundary conditions.

\subsection{Extended Bose-Hubbard model}
\label{extendedBH}

As promised in \sref{extendedBH_preview}, we first consider long-range interactions that are suddenly switched on (interaction quench) in an extended Bose-Hubbard model. This is realistic for engineered interactions in cold atom systems, e.g.~with Rydberg dressing.
For an initial coherent state, interactions will dephase the different Fock-state components, and coherence between neighbouring sites will be lost. At much longer timescales, the coherence may revive due to the limited number of relevant frequencies involved. Such collapse-and-revival sequences are well understood in the case of local interactions \cite{greiner:revival,kollath:quench}. 

Since the expression for the characteristic variance \bref{variance_final} was derived for ${\omega}_{nm}=0$, we shall consider $J=0$ first. We later demonstrate that for small nonzero $J$ the tools derived from \bref{variance_final} still offer useful guidance. For $J=0$, a state $|\bv{n}\rangle=|n_{1},n_{2},n_{3},...\rangle$ with exactly $n_{i}$ bosons on site $i$ is an eigenstate of the Hamiltonian and the time-evolution can be found analytically. Let us consider the case where the initial quantum state is a uniform coherent state on each site such as for a Bose-Einstein condensed gas $|\Psi(t=0)\rangle=|\phi,\phi,...\phi\rangle$, where $\phi\in {\mathbb C}$ is the BEC order parameter and $|\phi\rangle=\exp{(-|\phi|^2/2)}\sum_{n} \phi^n |n\rangle/\sqrt{n!}$. Thus $|\Psi(t=0) \rangle =\sum_{\bv{n}} C_{\bv{n}} |\bv{n}\rangle$, with $C_{\bv{n}}=\exp{(-| \bv\phi \cdot \bv\phi|/2)}\phi^{\sum n_{i}} /\sqrt{\prod n_{i}!}$. The state approximates the true state of a BEC in a lattice deeply in the superfluid regime. 
For our demonstration we compare Gauge-P simulations for $J=0$ with the analytical results and then consider various values of $J$ in Gauge-P only.

\begin{figure}
\centering
\epsfig{file={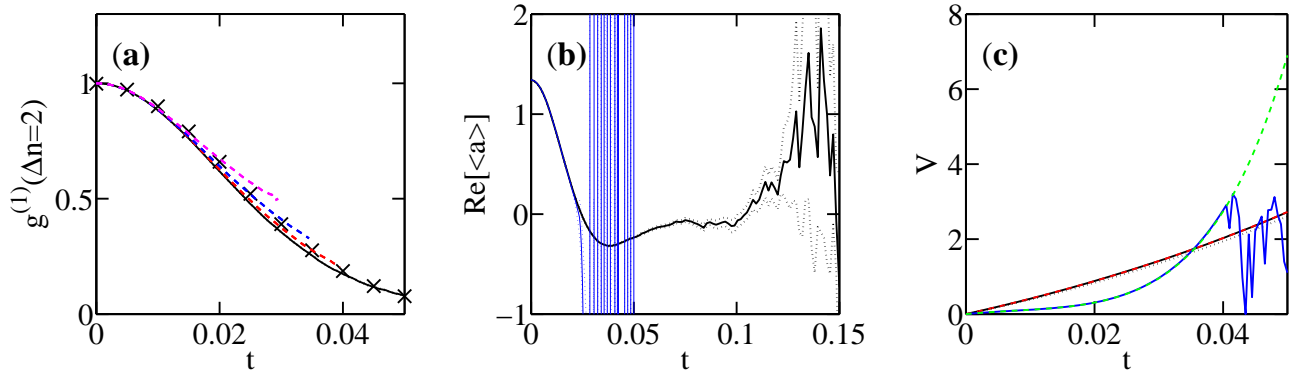},width=0.99\columnwidth} 
\caption{(colour online) (a) Dephasing of inter-site coherence after a sudden interaction quench in the extended Bose-Hubbard model, using the Gauge-P method: (solid black) $J=0$, (dashed red) $J=2$, (dashed black) $J=5$, (dashed magenta) $J=10$. All lines terminate at those times where simulations become intractable. ($\times$) Comparison with the analytical result for $J=0$. (b) The on-site mean field $\langle a \rangle$ for $J=0$ from Gauge-P is shown as solid black line, dotted lines indicating the sampling error. For comparison the blue-line shows a corresponding Positive-P simulation. Both use $10^5$ trajectories. (c) Characteristic variance ${\cal V}$ directly from the $J=0$ simulation, Gauge-P for a fixed $a=0.76$ (solid black),  Gauge-P with adaptive diffusion gauge as in \eref{adaptive_local_dg} (dotted black), Positive-P (blue). The corresponding analytical expression \eref{variance_final} (dashed red) and \eref{charact_variance_PosP_oft} (dashed green) give the exact variance since they are based on the complete Hamiltonian for this example. Note the different time-axis from (b). 
\label{BH_dephasing}}
\end{figure}

In \fref{BH_dephasing} we show the evolution of the inter-site coherence, $g^{(1)}(\Delta n)=\sum_{n} \langle a^\dagger_{n} a_{n+\Delta n} \rangle/\sqrt{n_{n}n_{n+\Delta n}}$, the on-site mean field $ \langle \hat{a}_n \rangle$ and the characteristic variance ${\cal V}$ for an interaction quench on $M=6$ sites spread on a periodic 1D interval of length $L=2$. The potential \eref{potential} is described by $C_{6}=-32$, $\epsilon=1$, $a=2$, $b=3$ and the initial state is $|\phi|^2$=$\rho\,dV=1.2$. $J$ is varied over $J=0,2,5,10$. 

To know in advance up to which times we can simulate, we determine $I_{2}$ from \eref{variance_localO5} using the potential \eref{potential}\footnote{$I_{1}$ is zero for our real coherent state initial conditions.}. Due to the small number of modes in this example, replacing the integral in $I_{2}$ by the corresponding discrete sum is more accurate, yielding $t<\sub{t}{\rm sim}=0.14$ using \eref{tsim3}.
A comparison with \frefp{BH_dephasing}{b} shows that this indeed describes the available time for the observable $\langle a_n \rangle$ quite well, for which the stochastic average is structurally similar to the characteristic variance \bref{charact_variance} underlying the estimate for $\sub{t}{sim}$. In contrast, the sampling of the inter-site coherence $g^{(1)}$  becomes intractable some time before $\sub{t}{\rm sim}$, around the maximal time shown in panel (a). This is common for higher order quantities. Being interested in $g^{(1)}$, we thus have chosen $t_{\rm opt}=0.05$ in the formula \eref{adaptive_local_dg}, which then gives $a=0.7$.

Finally, panel (c) numerically verifies \eref{variance_final}. It also shows that for the present model with conserved mean occupation of all the sites, the adaptive diffusion gauge $a(t,n)$ yields no significant improvement of the variance compared to a fixed gauge $a$. The probable reason is the constant mean mode occupation for this example. We expect that there exist cases with time-varying occupation where the adaptive gauge outperforms the fixed one.

Using the ungauged positive-P method we can only get results for a much shorter time shown in \frefp{BH_dephasing}{b}. The estimate \eref{tsimP} $t\lesssim 0.07$ is of the right magnitude, but a little high.

\subsection{Coulomb blockade}
\label{blockade}

We now consider the excitation of atoms within a homogenous 1d BEC to long-range interacting Rydberg states, as was outlined in \sref{rydberg_preview}. Establishing an interaction blockade within the excited state fraction will correspond to nontrivial correlation function dynamics. To see this within the accessible time interval up to $t_{\rm sim}$, let us consider an echo sequence as used in \cite{raitzsch:echo,Younge_raithel_echo}: Atoms are excited from the ground to the Rydberg state with a Rabi-frequency $\kappa$ during a time-interval $\tau/2$, then de-excited again using $\kappa\rightarrow-\kappa$ for a time $\tau/2$. This yields an additional time-scale $\tau$ which can be chosen such that
 if it is short enough, it can fit within $t_{\rm sim}$.
During this time, we are then interested in the time-evolution of the Rydberg-Rydberg correlation function
\begin{eqnarray}
g^{(2)}(r)&=\frac{\langle \Psi^{\dagger}_e(x)\Psi^{\dagger}_e(x+r)\Psi_e(x+r)\Psi_e(x) \rangle}{\rho_e(x) \rho_e(x+r)}, 
\label{g2def}
\end{eqnarray}
with $\rho_e(x) = \langle \Psi^{\dagger}_e(x)\Psi_e(x) \rangle$. Due to our homogeneous initial conditions, we can additionally average the expression \bref{g2def} over $x$.
The evolution equations are \eref{final_stoch_eqGaugeP} with the drift gauge \eref{driftg} and global diffusion gauge \eref{globalO}, adapted for two fields and the coupling $\kappa$ in \eref{Hamiltonian_Ryd}. 
Namely,
\begin{subequations}
\label{coulombeq}
\begin{eqnarray}
\movel
\pdiff{}{t}\gamma_{e,\mu}=i \sum_{\nu} \omega_{\mu \nu} \gamma_{e,\nu} \mp i\frac{\kappa}{2}\gamma_{g,\mu}+ i\sum_{\nu}\gamma_{e,\mu}W_{ \mu \nu}n'_{e,\nu} + \sqrt{i}\sum_{\nu\sigma} \gamma_{e,\mu}S_{\mu \nu}O_{\nu\sigma}\xi_{\sigma},
\\
\movel
\pdiff{}{t}\gamma_{g,\mu}=i \sum_{\nu} \omega_{\mu \nu} \gamma_{g,\nu} \mp i\frac{\kappa}{2}\gamma_{e,\mu},
\\
\movel
\pdiff{}{t}\Omega= i\sqrt{i}\Omega \sum_{\nu\sigma\lambda} n''_{e,\lambda}O_{\sigma\nu}S_{\lambda\sigma}\xi_{\nu}
\end{eqnarray}
\end{subequations}
with $\hbar=m=1$ and kinetic motion, if present, given by  $\omega=\frac{1}{2m}\nabla^2$. The sign $\mp$ refers to $-$ for $\alpha$ and $+$ for $\beta$ variables.

It would also be interesting to follow the entire development of the blockade-dip for larger times, and many-body Rabi oscillations in the saturated regime.
However, we found the time-scale for development of a complete blockade out of reach for all physical and gauge parameters tested here. 

\subsubsection{Echo sequence}
\label{echoseq}
\begin{figure}
\centering
\epsfig{file={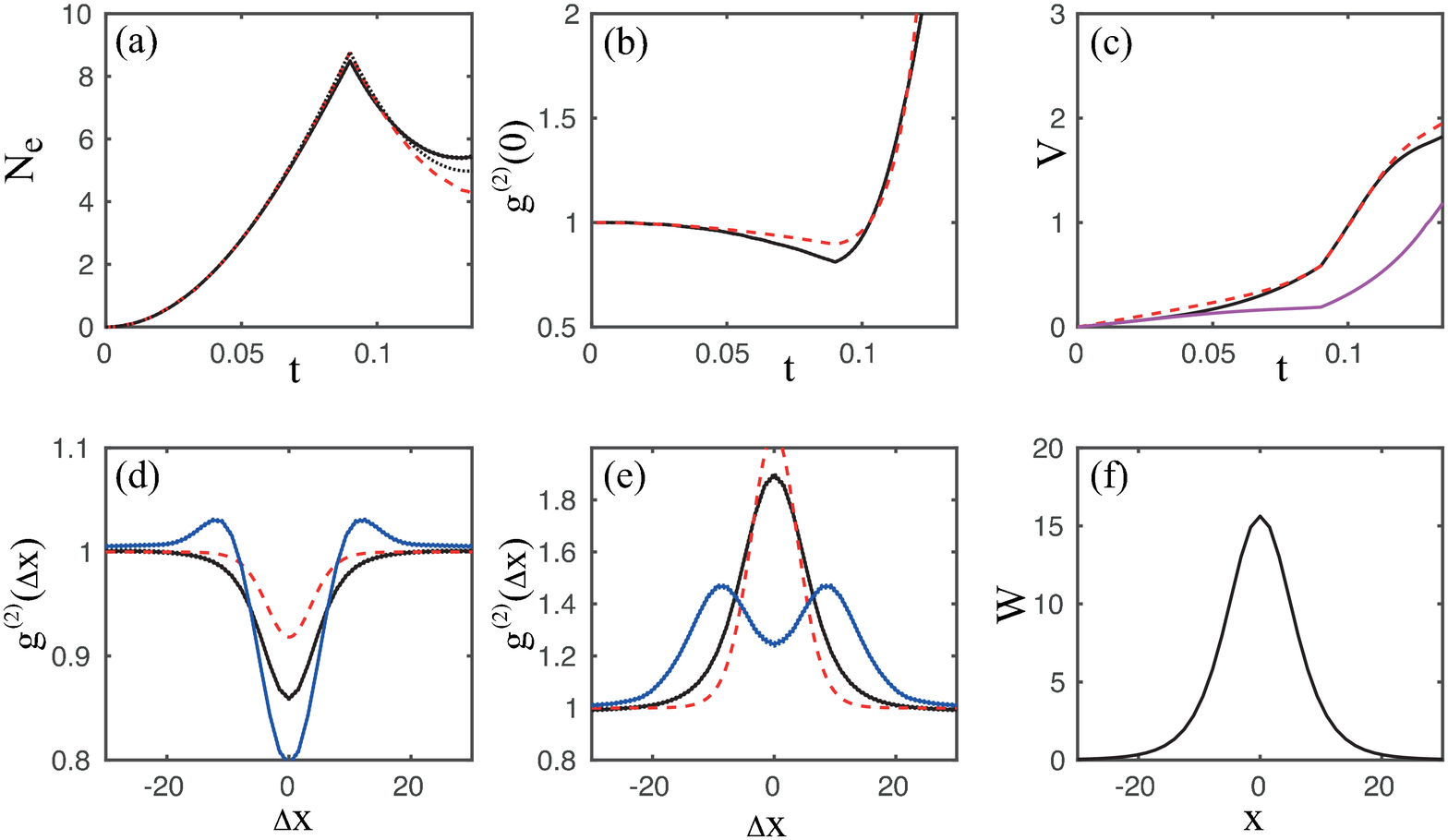},width=\columnwidth
} \\
\caption{(colour online) Rydberg excitation echo sequence as described in the text. (a) Excited state population $N_{e}$: (solid black) Gauge-P with adaptive gauge, (dotted black) mean-field, deviating only at the latest times, (dashed red) omega-expansion. (b) Rydberg-Rydberg correlation function at the origin $g^{(2)}(0,t)$, lines as in (a). (c) Characteristic variance ${\cal V}$, (solid black) adaptive gauge \eref{adaptive_local_dg}, (dashed red) fixed gauge $a=0.55$, (solid magenta) positive-P method. (d) Spatial correlations $g^{(2)}(\Delta x,t)$ at $t=0.08$ and (e) $t=0.12$, lines as in (a), additionally (solid blue), Gauge-P with atomic motion, i,e, $\omega=\frac{1}{2m}\nabla^2$. 
(f) Interaction potential \eref{potential}.
\label{echoplots}}
\end{figure}

Without interactions, after completing the echo sequence at time $\tau$, all the atoms would have returned to the ground-state. With interactions, in contrast, the induced dephasing makes this reversal incomplete. 
We model this scenario with  parameters $\tau=0.18$, $\kappa=3$, $C_6=-5.96\times10^{7}$, $\epsilon=12.5$ and initially $N=500$ atoms in state $\ket{g}$ between $x=-L$ and $x=L$ for $L=50$. These parameters are chosen for demonstration, to yield an excitation blockade that only allows a few atoms in state $\ket{e}$ within the domain $[-L,L]$.  Fig.~\ref{echoplots} shows some of the results. 
Except where indicated, we invoke the frozen gas approximation, in which the $\omega$ terms in \eref{coulombeq} are neglected. 

We find that, counterintuitively, the remnant excited state population after a portion of the de-exitation pulse develops {\em bunching} correlations as seen in \fref{echoplots}e. We confirmed this behaviour using the omega-expansion \cite{jovica:omegexp1,jovica:omegexp2}, which allows a qualitative calculation of the excited state population $N_{e}$ and correlations $g^{(2)}$ as an expansion in $\omega t$. The leading order results, shown in \fref{echoplots} for comparison, reveal qualitative agreement with the exact gauge-P with some quantitative differences. Preliminary calculations of the kind shown in \fref{echoplots} were the first sign of bunching correlations after an echo-sequence discovered by us. This prompted our more detailed investigation using exact diagonalisation for small Rydberg ensembles, published in \cite{wuester:correlations} and experimentally confirmed in \cite{Thaicharoen:correlcontrol}.


A good {\it a priori} estimate of gauge parameters is not so straightforward here like in the last section, because the guiding expressions that we have derived depend on the density of the interacting species $\rho_e=N_e/2L$, which is strongly time-dependent as seen in Fig.~\ref{echoplots}(a). 
First, using $N_e\approx 8$ from \frefp{echoplots}{a} as a worst-case estimate, we obtain for the Positive-P method $t_{sim}=0.21$ using \eref{tsimP}, while for Gauge-P we find $t_{sim}=0.32$ using \eref{tsim3}. These are roughly consistent with an empirical simulation time of $t_{\rm sim}=0.2$, but the numerics do not bear out the predicted modest advantage of the gauge-P here. We also note that the optimum fixed gauge found empirically, $a=0.55$, is somewhat smaller than that predicted by \eref{adaptive_local_dg}, which is $a_{\rm approx}=0.66$ for $t_{\rm opt}=0.135$. These differences do not come as a complete surprise, since our calculations neglect the strong time dependence of $n_e$.

While the frozen Rydberg gas dynamics presented so far can be alternatively calculated using exact diagonalisation \cite{wuester:correlations}, the quantum mechanical inclusion of motion typically renders that approach intractable, and would require the use of classical or quantum-classical methods \cite{cenap:motion,PhysRevLett.Amthor:mechanical:prl,Amthor:mechanical:pra,wuester:cradle,wuester:CI,leonhardt:switch,leonhardt:unconstrained,thaicharoen:trajectory_imaging,thaicharoen:dipole_trajectory_imaging,celistrino_teixeira:microwavespec_motion}.
Motion however poses no additional challenge to the Gauge-P method, which we now illustrate by dropping the frozen gas approximation and considering atomic motion as outlined in \sref{rydberg_preview}, with the full equations \eref{coulombeq}.
Simulated echo-sequences including motion are added to \fref{echoplots}. We used an exemplary mass of $m=4\times 10^{-3}$, chosen to produce a visible effect but not too short $\sub{t}{sim}$. As one can see, the change is significant.

There are many scenarios in which the frozen Rydberg gas approximation is not applicable such as\cite{cenap:motion,PhysRevLett.Amthor:mechanical:prl,Amthor:mechanical:pra,wuester:cradle,wuester:CI,leonhardt:switch,leonhardt:unconstrained,thaicharoen:trajectory_imaging,thaicharoen:trajectory_imaging,celistrino_teixeira:microwavespec_motion}. Other possible realisations of our long-range interacting model with relevant atomic motion could be
a two component polar BEC, where the long-range interactions between one of these components are dramatically enhanced with a Feshbach resonance \cite{Griesmaier:chromiumcondensate,Lahaye:ferrofluid} or a Rydberg dressed BEC \cite{laura:latticeclock,macri:latticedress,weibin:latticemotion:prl,zeiher:dressedlattice}. 

\section{Methods for very large systems}
\label{largeM}

The sets of stochastic differential equations \eref{ppequations} and \eref{gaugev} are usually quite straightforward to implement for small and medium size systems up to $M\sim\mathcal{O}(10-1000)$. 
However, for larger system two problems arise: The direct numerical implementation of the equations as written may become too inefficient, and the drift gauge may cease to be beneficial. We discuss both of these issues, and possible solutions, below.
\subsection{A more efficient treatment of the interaction for large systems}
\label{pp_altnoise}

For larger systems, the direct application of  \eref{ppequations} and \eref{gaugev} by summing over $m=1,\dots,M$ becomes a problem for two reasons: 
(i) integrating a single time step takes $\mathcal{O}(M^2)$ operations (a separate summation of $\sum_m W_{nm}\alpha_m\beta_m$ and $\sum_j [\sqrt{W}]_{nj}\xi_j$ for each lattice site $n$), and 
(ii) if $M$ gets really large, e.g., of the order of $M\gtrsim\mathcal{O}(10^5)$ in a three-dimensional system, calculating the $M\times M$ matrix $\sqrt{W}$ at the beginning of a calculation becomes time and memory consuming. 

However, the computational work could be substantially reduced by taking advantage of the fact that physical potentials $W(\mv{z})$ depend only on the vector difference $\mv{z}=\mv{x}-\mv{y}$. The overwhelming part of information in the two-index matrix $W_{nm}$ is redundant: $W_{nm}=W_{(n-m)\,{\rm mod}\,M,0}=W_{q0}$ with $q=(n-m)\,{\rm mod}\,M$, an index that embodies only the relative displacement on the numerical lattice.

Consider the discrete Fourier transform (DFT) of this translated interaction potential $W_{q0}=W(\mv{z}_q)$ on a numerical lattice with periodic boundary conditions:
\begin{equation}\label{FTW}
\wt{W}_{\wt{m}} = dV \sum_{q} W_{q0} e^{-i\mv{k}_{\wt{m}}\cdot\mv{z}_q} = dV e^{i\mv{k}_{\wt{m}}\cdot\mv{x}_m}\,\sum_{n} W_{nm} e^{-i\mv{k}_{\wt{m}}\cdot\mv{x}_n}.
\end{equation}
The last expression brings us back to the two-coordinate form $W_{nm}$ for all values of  $m$.
Here, $\mv{k}_{\wt{m}}$ are the standard wavevectors on the numerical lattice. Since the potential $W_{nm}$ is even, that is $W_{n,(n+d)\,{\rm mod}\,M} = W_{n,(n-d)\,{\rm mod}\,M}$ for any $d$ due to 1D periodicity (analogously in higher dimensions), the transform $\wt{W}$ will be real. The interaction drift term for $d\alpha_n/dt$ in \eref{ppequations} can then be written as an inverse DFT
\begin{equation}\label{driftMlogM}
-i\alpha_n\frac{1}{VdV}\sum_{\wt{m}} \wt{W}_{\wt{m}} \wt{n}_{\wt{m}} e^{i\mv{k}_{\wt{m}}\cdot\mv{x}_n},
\end{equation}
where $V=MdV$ is the total volume, tildes will indicate k-space, and
\begin{equation}
\wt{n}_{\wt{m}}(t) = dV \sum_{p} \alpha_p(t)\beta_p(t) e^{-i\mv{k}_{\wt{m}}\cdot\mv{x}_p}
\end{equation}
is the DFT of the density $n_p(t) = \alpha_p(t)\beta_p(t)$ for a given trajectory at a given time. 
This way, we can see that the deterministic evolution can be carried out by storing the interaction vector $\wt{W}_{\wt{m}}$ of just size $M$ from the beginning of a simulation, and by carrying out two DFTs per trajectory and time step to obtain $\wt{n}$. This has only $M\log M$ computational cost with ``fast Fourier transform'' methods. 

Remarkably, the noise terms can be dealt with in a related, though more involved way, that follows an approach reported in \cite{DeuarPhD-ranged}. Denoting the noise term for $d
\alpha_n/dt$ as 
$X_n = \sqrt{-i}\,\alpha_n\sum_j[\sqrt{W}]_{nj}\xi_j^{(1)}(t)$, the diffusion condition \eref{Dpp} requires simply that 
\begin{equation}\label{XX}
\overline{ X_n X_m } = -iW_{nm}\alpha_n\alpha_m
\end{equation}
 (equal times will be assumed in this section).
Then, if we rewrite the noise $X_n$ in terms of new scaled and Fourier-transformed noise quantities $\wt{Y}$ as
\begin{equation}\label{Xn0}
X_n = \sqrt{-i}\,\alpha_n \frac{1}{V} \sum_{\wt{p}} \wt{Y}_{\wt{p}} e^{i\mv{k}_{\wt{p}}\cdot\mv{x}_n},
\end{equation}
it can be shown by simple substitution that the condition
\begin{equation}\label{YY}
\overline{ \wt{Y}_{\wt{p}} \wt{Y}_{\wt{q}} } = V \wt{W}_{\wt{p}}\, \delta_{\wt{p},-\wt{q}}
\end{equation}
on the new noises is sufficient to satisfy the required diffusion condition \eref{XX}. If we are able to construct a noise field $\chi_{\wt{p}}$ with the somewhat atypical correlation properties 
\begin{equation}\label{ZZ}
\overline{  \chi_{\wt{p}} \chi_{\wt{q}} } = \delta_{\wt{p},-\wt{q}},
\end{equation}
then the choice $\wt{Y}_{\wt{p}} = \chi_{\wt{p}}\,\sqrt{V\,\wt{W}_{\wt{p}}}$ will satisfy all that is required. 
Such a noise can indeed be constructed, as is described in \aref{noiserewriting}.

With it, the noise term \eref{Xn0} can be written as an appropriate inverse DFT:
\begin{equation}\label{Xn}
X_n = \sqrt{\frac{-i}{V}}\,\alpha_n\, \sum_{\wt{p}} \chi_{\wt{p}}(t)\, \sqrt{\wt{W}_{\wt{p}}}\, e^{i\mv{k}_{\wt{p}}\cdot\mv{x}_n}.
\end{equation}
This is also implementable with computational cost $M\log M$ per time step. The linear term $\sum_m\omega_{nm}\alpha_m$ does not usually constitute an efficiency problem, so we leave it as is.
Applying the above, using independent noise fields $\chi^{(1)}$ and $\chi^{(2)}$, the final more efficient equations in the positive-P representation are
\begin{subequations}\label{ppequationsMlogM}
\begin{eqnarray}
\frac{d\alpha_n}{dt}&= -&i\sum_m \omega_{nm}\alpha_m -i\alpha_n \frac{1}{dV}\sum_{\wt{m}} \left[ \frac{\wt{W}_{\wt{m}} \wt{n}_{\wt{m}}}{V} -\frac{\sqrt{i dV}}{M}\, \chi^{(1)}_{\wt{m}}(t)\, \sqrt{\wt{W}_{\wt{m}}}\right] e^{i\mv{k}_{\wt{m}}\cdot\mv{x}_n},\qquad\qquad\\
\frac{d\beta_n}{dt}&=& i\sum_m \omega_{nm}\beta_m -i\beta_n \frac{1}{dV}\sum_{\wt{m}} \left[ \frac{\wt{W}_{\wt{m}} \wt{n}_{\wt{m}}}{V} -i\frac{\sqrt{i dV}}{M}\, \chi^{(2)}_{\wt{m}}(t)\, \sqrt{\wt{W}_{\wt{m}}}\right] e^{i\mv{k}_{\wt{m}}\cdot\mv{x}_n}.\qquad\qquad
\end{eqnarray}
\end{subequations}
%

\subsection{Drift gauges and system size}

Looking at the simulation time estimate for the gauge-P distribution in the contact-interacting case \eref{tsim3c}, one sees that after intensive quantities like $g$ and $\rho$ have been factored out, a disadvantageous reduction with the system size $M$ remains.  This is a feature that has been also noted previously. It arises because the single weight variable collects noise from the entire system\cite{deuar:jphysAII}, while the limit ${\cal V}\lesssim10$ remains unforgiving as system size grows.  In contrast, positive-P or only diffusion gauged calculations do not show this behaviour. For this reason, though they are consistently less advantageous for single modes, they may be better when the system size becomes large. 

 \subsubsection{Diffusion gauge only}
\label{diffusionG}
To address these issues, we first evaluate how a global diffusion gauge \eref{globalO} performs in long-range interacting systems without any drift gauges.  
We proceed the same way as in \sref{gaugechoice} and \ref{tsims} and \cite{deuar:jphysAII}. The variance \eref{variance_finalP} is expanded to third order in $t$, with gauge \eref{globalO}, and deterministic initial conditions $C^{(0)}=\widetilde{C}^{(0)}=0$.
The result is:
\begin{eqnarray}
\movel
{\cal V}^{(P)}(t)={\cal V}(0) + \frac{tU_0}{2}\,\cosh2a-\frac{t^2}{2}I^{(P)}_{1} +\frac{t^{3}}{3}\,e^{-2a}\,I^{(P)}_{2}
\label{variance_finalD}
\end{eqnarray}
with the same integrals as in \eref{charact_variance_PosP_oft_expanded}.
The $I_1^{(P)}$ term is irrelevant as it does not depend on $a$. The optimum gauge then is readily shown to be simply 
\begin{eqnarray}
\movel\label{aoptD1}
a_{\rm opt} \approx \frac{1}{4}\log\left(\frac{4t_{\rm opt}^2I_2^{(P)}}{3U_0}+1\right),
\end{eqnarray}
in agreement with known single-mode results. 
From the estimates \eref{IestP}, we see that this is largely independent of $M$, as hoped. 

The simulation time, requiring ${\cal V}(t_{\rm sim})=10$ can be found in the usual two regimes:
The noise amplification dominated regime, when $n$  and $t_{\rm sim}$ are large enough, has
\begin{eqnarray}
\movel
t_{\rm sim}\approx \frac{{\cal O}(4)}{\left[U_0I^{(P)}_2\right]^{1/4}}. 
\label{tsimD}
\end{eqnarray}
This is a different and more advantageous scaling as compared to the plain positive-P \eref{tsimP}. 
The diffusion gauged value of $t_{\rm sim}$ is longer when
\begin{equation}\label{I2cond}
I_2^{(P)} \gtrsim 0.03\, U_0^3.
\end{equation}
Notably, $I_2^{(P)}$ is an integral involving third powers of the potential, so \eref{I2cond} is primarily a condition on the density, requiring it to be sufficiently high. Substituting the estimates \eref{IestP}, one has 
$n_0\gtrsim0.2$ for the lattice occupations. This result does not depend on system size.
The direct-noise dominated regime has the same estimate \eref{tsimP1} as in the other cases.  

\subsubsection{Diffusion gauges vs drift gauges}
\label{vs}
With the different $t_{\rm sim}$ of drift and diffusion-only gauged calculations, a pertinent question is when should the drift gauge be added to the diffusion gauge?
Comparing \eref{tsim3} and \eref{tsimD}, diffusion only calculations last longer when 
\begin{equation}\label{Icomparison}
I_2^{(P)} \lesssim \frac{U_0}{16}\,I_2.
\end{equation}
For contact interactions in a uniform system when substituting \eref{Iest} and \eref{IestP} this reduces to the surprisingly simple expression 
\begin{equation}
M\gtrsim 16.
\label{M16}
\end{equation}
For long range interactions, condition \eref{Icomparison} should be used instead of \eref{M16}. Also there it is clear, however, that for large enough systems, say $M\gg100$, the diffusion gauge only approach will be better. 
 
The simulations of \sref{blockade} were an example of a sufficiently large system, with $M=64$, $I_2^{(P)}\approx 2500$, $U_0=15.624$, and $I_2\approx1500$, where drift gauges cease to offer a huge advantage.

\section{General many-mode diffusion gauges}
\label{nonlocal_diff}
We have seen so far, that the global gauge ansatz \bref{globalO} can already provide significant advantages as well as technical challenges in the implementation.
However, it only exploits a small fraction of the degrees of freedom for the many-mody case that are inherent in the full $2M\times 2M$ matrix $\mvv{O}$. In this section we present our initial explorations of \emph{non-local many-mode} Gauges. These are intended as starting point for future research, beyond the scope of this article.

For many-mode problems with contact potentials $W= W_0 \id$ as considered in previous work, a \emph{local diffusion gauge}
\begin{eqnarray}
\label{globalOa}
\mvv{O}=\left( 
\begin{array}{cc}
\mbox{diag}(\cosh a_{n}) & -i \:\mbox{diag}(\sinh a_{n}) \\
i\: \mbox{diag}(\sinh a_{n}) & \mbox{diag}(\cosh a_{n})
\end{array} 
\right)
\end{eqnarray}
is a feasible choice that already goes beyond \bref{globalO}. Since $\mvv{O}$ and $\mvv{S}$ in \eref{final_stoch_eqOmega} commute in this case, each gauge parameter $a_{m}$ can be individually associated with the mode $m$. For sufficiently small inter-mode coupling, one can then use an optimized gauge $a_{m}(t)$ based on single-mode results that depend on the stochastic occupation $n_{m}$
of this particular mode $m$. Often this gives good results~\cite{deuar:thesis}.

However, for a non-local potential $W$, the matrices $\mvv{O}$ and $\mvv{S}$ no longer commute and hence the $a_{n}$ would have to be viewed as gauge parameter of the noise source $\xi_{n}$. In that case, it is not clear to what degree a naive application of single-mode results is still possible. In the remainder of this section we describe two approaches aimed at harnessing more of the gauge freedom of $\mvv{O}$: an analytical one and a numerical one. 

\subsection{Analytical nonlocal diffusion gauges}
\label{analytical_nonlocal_diff}

Finding the matrix  $\mvv{O}$ which globally minimizes the variance for a given interaction potential and simulation time is a difficult task, analytically and numerically. Here, we investigate the ansatz
\begin{eqnarray}
\mvv{O}=\left( 
\begin{array}{cc}
\cosh A &-i \sinh A \\
i \sinh A & \cosh A
\end{array} 
\right),
\label{nonlocalO}
\end{eqnarray}
where the hyperbolic functions are defined in terms of matrix exponentials, and $A$ is a symmetric $M\times M$ matrix. The matrix $\mvv{O}$ is orthogonal for all choices of $A$ and reduces to \bref{globalO} for $A= a \id$, with scalar $a$.

Now let us consider the subset of nonlocal interactions for which $W=U$, i.e. $\sqrt{W}=\sqrt{W}^{\dagger}=\mbox{Re}[\sqrt{W}]$. 
We take $A$ real and the starting densities deterministic: $\bv{n}(0)=\overline{\bv{n}(0)}$.
Then, inserting $\mvv{O}$ into the variance expression \eref{variance_final} and simplifying, we obtain the following for short times:
\begin{eqnarray}
\movel
{\cal V}=\frac{t}{2M}\mbox{Tr}\left[W\cosh(2 A)\right]  + t\,\bv{n}^{\prime\prime T}(0)\sqrt{{W}} \exp(-2 A)\sqrt{{W}}\bv{n}''(0) 
\CR
+ \frac{t^2}{2}\mbox{Tr}\left[\sqrt{{W}} \exp(-2 A)\sqrt{{W}} N' \sqrt{{W}} \exp(-2 A)\sqrt{{W}} N' + (N' \rightarrow N'' )\right].
\label{variance_nonlocalO}
\end{eqnarray}
We used the matrices 
\begin{eqnarray}
N'={\rm diag}[\bv{n}'(0)],\qquad N''={\rm diag}[\bv{n}''(0)].
\label{NiNii}
\end{eqnarray}
A minimum of the scalar quantity ${\cal V}$ in the space of matrices $A$ is given by the solution of $\partial {\cal V}/\partial A_{ij}=0$.

As described in more detail in \aref{nonlocalDG}, if $\bv{n}$ is homogeneous and real, we can obtain the matrix equation
\begin{eqnarray}
 {W}= {W} \exp(-4 A)  + 4 Mt \left[ \sqrt{{W}} N' {W} N' \sqrt{{W}} \right] \exp(-6 A) .
\label{nonlocalAmatrixequation}
\end{eqnarray}
Equation \eref{nonlocalAmatrixequation} has the same structure as the single mode version in~\cite{deuar:jphysAII}. Hence, following~\cite{deuar:jphysAII}, we can solve this equation separately for cases where the $\exp(-4 A)$ term or $\exp(-6 A)$ term on the right hand side is dominant, and then suggest a simple interpolation equation:
\begin{eqnarray}
A=\frac{1}{6}\log\bigg\{ 4 M t_{\rm opt} \sqrt{{W}}^{-1}N' {W}N'\sqrt{{W}}
+ \id \bigg\},
\label{optimalA}
\end{eqnarray}
where $\log$ denotes a matrix logarithm. 
We present \eref{optimalA} for completeness, as for the cases explored, it did not prove more useful than the much simpler adaptive global gauge of \sref{globalDG}. 

We defer an initial test of the performance of \bref{optimalA} to the next section.

\subsection{Numerical nonlocal diffusion gauges}
\label{numerical_nonlocal_diff}

Another route by which non-local diffusion gauges might offer an advantage over local adaptive ones, is the numerical minimization of  \eref{variance_final} with respect to an arbitrary orthogonal matrix $O$ for given initial mode occupations $n_m(0)$, $t$ and ${W}$. We carried out such a minimization for the same potential as in \sref{echoseq} with spatial atom-density $n(x)=\bar{n}\exp[-x^2/2/\sigma^2]$, $\sigma=10$, $\bar{n}=0.5$ and $t_{\rm opt}=0.15$. 
A small number of modes $M=16$ covering the range $x\in[-L,L]$ with $L=50$ allowed reasonably fast optimization. In practice the degree of freedom for the conjugate gradient optimization routine \cite{blair:thesis} was a real, anti-symmetric matrix $g$ such that $O=\exp(g)$ as in \eref{globalO}.

\begin{figure}
\centering
\epsfig{file={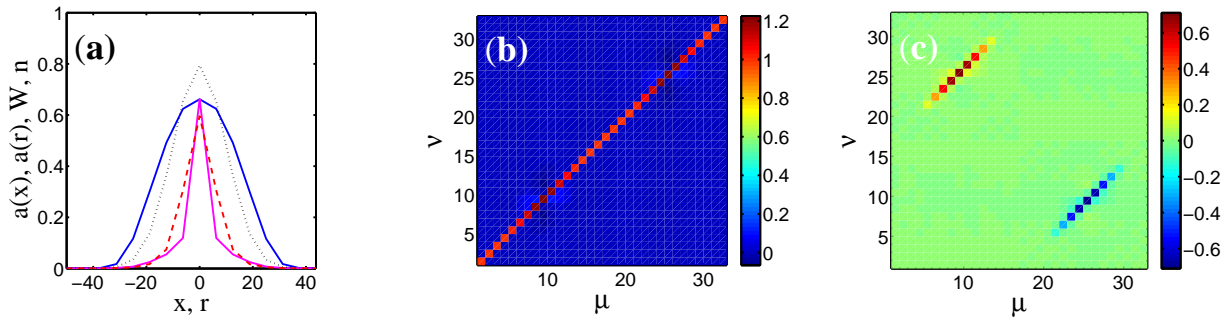},width=0.99\columnwidth} 
\caption{(colour online) 
A numerical optimization of the non-local diffusion gauge \eref{nonlocalO}.
(a) Physical conditions are given by the potential $W(r)$ (red-dashed) and density $n(x)$ (black dotted) described in the text, shown here in arbitrary units. We then show the resulting spatial and non-local variations of the diffusion gauge using $a(x)$ (blue solid) and $[a(r)]$ (magenta solid), respectively, as described in the text.
(b) Re[$O_{\mu\nu}$], (c)  Im[$O_{\mu\nu}$].
The initial condition (i) described in the text was used.
\label{fig3}}
\end{figure}
We employed a variety of different initial conditions: (i) local global gauge \bref{globalO} with $a$ guessed, (ii) non-local gauge \bref{nonlocalO} with $A_{kl}=$const, (iii) $\mvv{O}=\exp(g)$ with random $g_{kl}$, (iv) $g=0$, (v) the analytical solution \eref{optimalA}. For comparison with local diffusion gauges \eref{globalOa}, we extracted from the optimized $\mvv{O}$ a local diffusion gauge parameter $a(x_\mu)=\sinh^{-1}[\mbox{Im}(\mvv{O}_{M+\mu,\mu})]$ for $1\leq \mu \leq M$, as well as the non-local shape $a(r_\mu)=\sinh^{-1}[\mbox{Im}(\mvv{O}_{M,\mu})]$, both shown in \fref{fig3} for the initial condition (i). For comparison, a pure local diffusion form \eref{globalOa} would contain nonzero elements only exactly on the dominant diagonal features described by $g(x)$ as seen in \frefp{fig3}{b--c}, while the magenta line for $g(r)$ in \frefp{fig3}{a} would only have the one nonzero element at $r=0$.

All initial conditions led to quite similar results as shown, with a significant spatial dependence of the gauge parameter $a(x)$ due to the inhomogeneous density $n(x)$, and small non-local tails visible in $a(r)$. The overall decrease in variance of \eref{variance_final} compared to a local diffusion gauge was by a factor of $2.3$ in the case shown, with weak non-locality. We also used the numerical optimisation scheme to confirm that \bref{optimalA} is indeed an optimal choice if $n(x)$ is uniform and times are short. Even for the inhomogeneous test case described above it already outperforms the global gauge by a factor of $2.1$, close to the performance of the completely optimized result. The expression \bref{optimalA} also roughly captures the off-diagonal shape of gauge parameters found by the numerical optimisation in those cases.
 
From this reduction of the characteristic variance ${\cal V}$, one can make one important general conclusion -- that there is indeed some potential for gains from a nonlocal optimization of the gauge. Since we just skimmed the surface of gauges more general than \eref{globalO}, this indicates an extensive future route for possible improvement in our simulation methods, beyond the scope of the present article. 

\section{Summary}

We have provided the necessary apparatus for carrying out stochastic positive-P and Gauge-P simulations for systems of long-range interacting bosons. 
Our focus has been on directions relevant for ultracold gases such as Rydberg and dipolar atoms, as this is presently a burgeoning field of experimental and theoretical investigation.  
 A summary of our main analytical results is given in \tref{mainresults} for easy reference.

Our most important results are the dynamical equations with a long-range potential, the estimates for the available simulation times $t_{\rm sim}$, and simple formulae for optimal diffusion gauge parameters $a_{\rm opt}$. The simulation times reveal whether a particular problem is amenable to calculation using this method, while the gauge formulae give a straightforward means of improving simulation times beyond those directly available from the positive-P method. The underlying general expressions for the noise evolution (the characteristic variances ${\cal V}$) constitute a good starting point for future investigations, including the development of truly non-local diffusion gauges. The results of Sec.~\ref{numerical_nonlocal_diff} confirm that nonlocal gauges can offer advantages in principle.

Comparing with the previously known results for contact interactions, we can make several statements: stochastic gauges continue to improve simulation times. For this we use a global gauge that is optimised with the help of nonlocal integrals such as $I_2$ and $I_2^{(P)}$. The resulting  expressions for the optimal gauge, simulation time, and characteristic variance that we have derived bear structural similarity to the contact interacting case, and reduce to them in the appropriate limit. For large systems with long-range interaction, special care must be taken to avoid a quadratic scaling with system size and memory problems, but remarkably, this can be achieved using Fourier transformed interactions and noise fields. This technique makes simulations with, say, $M=10^5-10^6$ possible, which would otherwise be well out of reach. 

Furthermore, we have confirmed here a suspicion that arose already for contact-interacting gases, that very large systems with $\gtrsim{\cal O}(100)$ modes have different stochastic gauge properties than small to medium systems. Drift gauges become inefficient for large systems because of the accumulation of noise from all modes into the one global weight $\Omega$. In that case, diffusion gauges should be used, and continue to provide improvement over the plain positive-P treatment for many cases. 

Finally, we have illustrated our techniques by modeling an interaction quench in the extended Bose-Hubbard model and the excitation of Rydberg states in a Bose-Einstein condensate, using an echo sequence. Notably, the inclusion of atomic motion in such a calculation introduces no additional significant complications here, in contrast to many other methods used to study Rydberg excitations.

\begin{table}[htb]
\begin{tabular}{|l|l|l|}
\cline{1-3}
\Big. 
{\bf Result}      & {\bf positive-P} & {\bf gauge-P}   \\
\cline{1-3}
Evolution equations     & basic \bref{ppequations}  &     general gauge-P \bref{final_stoch_eqGaugeP}\\
for nonlocal interactions:	  &  large system \bref{ppequationsMlogM}  &  standard drift gauge \bref{driftg}\\
\cline{1-3}
Characteristic variances:     &    \bref{charact_variance_PosP_oft}  &   all gauges \bref{variance_final}\\
        					  &  									 &  diffusion gauges only  \bref{variance_finalP}\\
\cline{1-3}
Optimized global gauges: & 						& all gauges \bref{adaptive_local_dg}\\
					&						&diffusion gauge only \bref{aoptD1} \\
\cline{1-3}
Simulation time estimate: & \bref{tsimP}  & all gauges \bref{tsim3}\\	
			&					 & diffusion gauge only \bref{tsimD}\\
			&				& gauge comparison \bref{Icomparison}\\
\cline{1-3}	
\end{tabular}
\caption{Summary of main analytic results.
\label{mainresults}}
\end{table}
%
 \section*{Acknowledgements}

It is a pleasure to thank Andr{\'e} Ribeiro de Carvalho, Karen Kheruntsyan and Murray Olsen for comments. We also gratefully acknowledge useful discussions with Cenap Ates, Thomas Pohl and Jovica Stanojevic. P.D. acknowledges support from the National Science Centre (Poland) grant No. 2012/07/E/ST2/01389.

\appendix
\section{Stratonovich corrections}
\label{stratocorrapp}

The evolution equations  \eref{final_stoch_eqalphabeta}--\eref{final_stoch_eqOmega} are to be interpreted in It{\^o} stochastic calculus. Most numerical implementations that go beyond simple and inefficient first-order methods require the use of the Stratonovich form. This includes the adaptive 8th/9th order Runge-Kutta method \cite{wilkie:adaptivestoch} employed for all simulations in this article using the XMDS package used here \cite{xmds1} (see \cite{XMDS,xmdspaper} for the most recent version). 
For equations with variable-dependent noise such as the ones derived here, the Stratonovich correction can be complicated, especially when the gauges themselves are a function of the variables. 
Hence, it is of utility to provide the end results here. 

If one decomposes the variables in $\overline{\bv{v}}=(\bgamma,\Omega)$ into all the independent variables (either ${\rm Re}[v]_{\mu}$ and ${\rm Im}[v]_{\mu}$ or $v$ and $v^*$), labeling them them as $u_j$ with $j=1,\dots2(2M+1)$, then 
one can apply the usual formula~\cite{book:qn,book:stochmeth} for the Stratonovich correction:
\begin{eqnarray}
S_{u_j}=-\frac{1}{2}\sum_{\mu}\sum_{k=1}^{2(2M+1)} B_{k\mu}\pdiff{}{u_{k}}B_{j\mu}
\label{stratocorr}
\end{eqnarray}
for  a real random process with It\^o equation 
\begin{equation}\label{stochasticdifferential}
\frac{du_j}{dt}=A_j(\overline{\bv{v}}) + \sum_{\mu} B_{j\mu}(\overline{\bv{v}}) \xi_{\mu}.
\end{equation}
The Stratonovich form of the equation is then 
$\frac{du_j}{dt}=A_j(\overline{\bv{v}}) + \sum_{\mu} B_{j\mu}(\overline{\bv{v}}) \xi_{\mu} + S_{u_j}(\overline{\bv{v}})$.

\subsection{Uniform diffusion gauge}
\label{conststratocorrapp}

With a diffusion gauge $\mvv{O}$ independent of the variables, but allowing arbitrary drift gauges $\bvv{f}$, we obtain: 
\begin{eqnarray}
S_{\gamma_{\mu}}&=&-\frac{i}{2}\gamma_{\mu} W_{\mu\mu}\label{Sab}
\\
S_{\Omega}&=&
-\frac{i}{2}\Omega \left[
\bvv{f}^T\, \mvv{W}\, \bvv{f} 
+ \sum_{\nu} \gamma_{\nu} \left[\mvv{W}\pdiff{\bvv{f}}{\gamma_{\nu}}\right]_{\nu} 
- i \sum_{\nu} \gamma^{*}_{\nu}\left[\mvv{S}\mvv{O}\mvv{O}^{\dagger}\mvv{S}^{\dagger} \pdiff{\bvv{f}^*}{\gamma_{\nu}}\right]^{*}_{\nu} \right].\qquad\qquad
\label{stratocorr_final}
\end{eqnarray}
The notation $\left[ \bvv{a} \right]_{\nu}$ is used for the $\nu$-component of the vector inside the brackets.

If we specify the drift gauge as $\bvv{f}=i\bvv{n}''$ from \eref{driftg} as we have used throughout this paper, one obtains
\begin{eqnarray}
S_{\Omega}=\frac{\Omega}{4}\sum_{\nu} n^{*}_{\nu} \left[\mvv{S}\mvv{O}\mvv{O}^{\dagger}\mvv{S}^{\dagger}\mvv{F}\right]^{*}_{\nu\nu}
\label{stratocorr_finalest}
\end{eqnarray}
with $\mvv{F}$ defined in \eref{Fdef}. 
A case relevant for this paper is the uniform diffusion gauge \eref{globalOa} using a real $\bv{a}$, where one has:
\begin{eqnarray}
S_{\Omega}=\frac{\Omega}{2} \left(\sum_n n_n \sum_m  [\sqrt{W}]_{nm}e^{-2a_m}[\sqrt{W}^{\dagger}]_{mn}\right)^*,
\label{stratocorr_const}
\end{eqnarray}
which reduces to 
\begin{eqnarray}
S_{\Omega}=\frac{\Omega}{2} \sum_n n^*_n e^{-2a}U_0 = \frac{\Omega U_0e^{-2a}}{2}\int d\mv{x}\, \rho(\,\mv{x})^*
\label{stratocorr_global1}
\end{eqnarray}
for the global gauge \eref{globalO}. The real rectified potential \eref{Udef} appears again.
The case of contact interactions is recovered when $W_{nm}=g\delta_{nm}/dV$. An ungauged positive-P simulation, or one with diffusion gauges that do not depend on the variables has only the \eref{Sab} corrections.

\subsection{Adaptive gauge Stratonovich correction}
\label{adaptivestratocorrapp}

In the case of an adaptive diffusion gauge the Stratonovich correction becomes significantly more complicated, since the derivatives in \eref{stratocorr} now also act on the gauge parameter $a$. 
Taking the global form \eref{globalO} but with the variable and time dependence of \eref{adaptive_explicit}, after much algebra, one obtains:
\begin{subequations}
\label{manymode_adaptive_sc}
\begin{eqnarray}
\movel
S_{\alpha(\mv{x})}&=\:\:\frac{{i}}{2} W_0\,{\alpha(\mv{x})} + i \frac{{\alpha(\mv{x})}\,e^{-6a}}{{U_0}}\Bigg\{\frac{{(t_{\rm fin}-t)}}{3}\left[I_{3}(\mv{x}) - ie^{- 2a}I_{4}(\mv{x})^* \right]
\CR
&\hspace*{1cm}+\frac{1}{2}\sqrt{1 + \frac{4 I_{1}}{W_0} }\left(-iI_{5}(\mv{x}) + e^{- 2a}I_{6}(\mv{x})^* \right)
\Bigg\},
\\
\movel
S_{\beta(\mv{x})}&= -\frac{{i}}{2}W_0\,{\beta(\mv{x})}  - i \frac{{\beta(\mv{x})} \,e^{-6a}}{{U_0}}\Bigg\{\frac{{(t_{\rm fin}-t)}}{3}\left[I_{3}(\mv{x}) +ie^{- 2a}I_{4}(\mv{x})^* \right]
\CR
&\hspace*{1cm}-\frac{1}{2}\sqrt{1 + \frac{4 I_{1}}{W_0} }\left[(I_{5}(\mv{x}) + e^{- 2a}I_{6}(\mv{x})^* \right)
\Bigg\},
\\
\movel
S_{\Omega}&= \frac{\Omega U_0}{2}e^{-2a}\int d\mv{x} \:\rho^{*}(\mv{x}) 
\CR
&- \frac{\Omega\,e^{-8a}}{U_0}\int d\mv{x}\, \rho''(\mv{x}) \left\{
-\frac{2i(t_{\rm fin}-t)}{3}\, I_{4}(\mv{x})^* 
+ \sqrt{1 + \frac{4 I_{1}}{W_0} }\, I_{6}(\mv{x})^*
\right\}\quad
\end{eqnarray}
\end{subequations}
\begin{subequations}
\begin{eqnarray}
\movel
I_{3}(\mv{x})= \int d\mv{z} d\mv{y}\: \rho(\mv{y})\rho(\mv{z})^*\: U(\mv{y}-\mv{z})^{2} W(\mv{x}-\mv{y}),
\\
\movel
I_{4}(\mv{x})= \int d\mv{z} d\mv{y}\: \rho(\mv{y})\rho(\mv{z})^*\: U(\mv{y}-\mv{z})^{2} U(\mv{x}-\mv{y}),
\\
\movel
I_{5}(\mv{x})= \int d\mv{z} d\mv{y}\: \rho(\mv{y})\rho''(\mv{z})\: U(\mv{y}-\mv{z}) W(\mv{x}-\mv{y}),
\\
\movel
I_{6}(\mv{x})= \int d\mv{z} d\mv{y}\: \rho(\mv{y})\rho''(\mv{z})\: U(\mv{y}-\mv{z}) U(\mv{x}-\mv{y}).
\end{eqnarray}
\end{subequations}
%

\section{Characteristic variance: with drift gauge}
\label{charvarapp}
 
\subsection{Logarithmic variables} 
To estimate \eref{charact_variance}, a first step is  to turn the stochastic equations (differentials) \eref{final_stoch_eqalphabeta}-\eref{final_stoch_eqOmega} into equations for $\log{|\Omega\gamma_{\mu}|}$. 
We use the usual multivariate It{\^o} formula~\cite{book:stochmeth,arnold:sdes} using the independent variables $\overline{\bv{v}}$ like in \eref{stochasticdifferential}, 
to give
\begin{eqnarray}
\movel
\frac{d}{dt}\log{(\Omega \gamma_{\mu})}&=&
\frac{i}{\gamma_{\mu}}\left[\mvv{\omega}\bvv{\gamma}\,\right]_{\mu}
+i\left[\mvv{W}(\bvv{n}-\bvv{f})\right]_{\mu}
-\frac{i}{2}\left[ W_0 + \bvv{f}^T\mvv{W}\bvv{f}\,\right]
\CR
&&+\sqrt{i}\left[\mvv{S}\mvv{O}\bvv{\xi}\,\right]_{\mu}
+ \sqrt{i}\bvv{f}^T\mvv{S}\mvv{O}\bvv{\xi}.
\label{dlog}
\end{eqnarray}
From this point on, we will neglect kinetic terms, setting $\mvv{\omega}=0$, as explained in section \ref{noiseevol_PosP}. 
Formally, \eref{dlog} can be integrated over time, with the result 
\begin{eqnarray}
\movel
\log{[\Omega(t) \gamma_{\mu}(t)]}
=
\log{[\Omega(0) \gamma_{\mu}(0)]}
 + i\left[\,\mvv{W}\int_{0}^{t}(\bvv{n}(s) -\bvv{f}(s))\,ds\right]_{\mu} - i\frac{W_0}{2}t
\CR
-\frac{i}{2}\int_{0}^{t} \bvv{f}^T(s)\,\mvv{W}\bvv{f}(s)\, ds  
+ \sqrt{i} \left[\mvv{S}\mvv{O}\int_{0}^{t}\bvv{\xi}(s)\,ds \right]_{\mu}
\CR
+ \sqrt{i} \int_{0}^{t} \bvv{f}^T(s)\,\mvv{S}\mvv{O} \bvv{\xi}(s).
\label{logt}
\end{eqnarray}
Next, we move from $\log{(\Omega \gamma_{\mu})}$ to $\log{(|\Omega \gamma_{\mu}|)}$. It is helpful to  specify the form of $\bv{f}$ at this point, as the usual $\bvv{f}=i\bvv{n}''$. 
This has the convenient feature that the deterministic parts of $\log{(\Omega \gamma_{\mu})}$ are imaginary and do not contribute to $\log{|\Omega \gamma_{\mu}|}$ since $\log|z|={\rm Re}[\log z]$. This is likely an important factor why this is a useful gauge.
With this choice of $\bvv{f}$, we arrive at
\begin{eqnarray}
\movel
\log{[|\Omega(t) \gamma_{\mu}(t)|]}
=
\log{[|\Omega(0) \gamma_{\mu}(0)|]} + \frac{\sqrt{i}}{2} \sum_{\nu}\left[\mvv{S}\mvv{O} -i\mvv{S}^*\mvv{O}^{*}\right]_{\mu\nu}\int_{0}^{t}\xi_{\nu}(s)ds 
\CR
+  \frac{i\sqrt{i}}{2}\sum_{\nu\lambda} \left[\mvv{S}\mvv{O} +i\mvv{S}^{*}\mvv{O}^{*}\right]_{\lambda\nu}  \int_{0}^{t}n''_{\lambda}(s)\xi_{\nu}(s).
\label{logmodt}
\end{eqnarray}
As previously, the indices below square brackets denote the elements of the enclosed matrix/vector.

\subsection{Variance assembly} 

The mean of \eref{logmodt} is simply $\overline{\log{[|\Omega(t) \gamma_{\mu}(t)|]} } = \overline{\log{[|\Omega(0) \gamma_{\mu}(0)|]}}$, since $\overline{ \int_{0}^{t}G(s)\xi_{\sigma}(s)}=0$ for any non-anticipating function $G$~\cite{book:stochmeth}. The remaining averaging at $t=0$ is over initial state noise, which is independent of any later dynamical noises $\xi$. The
mean of the square is
\begin{subequations}
\label{meanloglog}
\begin{eqnarray}
\movel
\movel
\overline{\log{[|\Omega(t) \gamma_{\mu}(t)|]}^2}
=
\overline{\log{[|\Omega(0) \gamma_{\mu}(0)|]}^{2} }\mbox{\hspace{0.6cm}}
\CR
\movel
+ \frac{i}{4} \sum_{\nu\nu'}\left[\mvv{S}\mvv{O}-i\mvv{S}^{*}\mvv{O}^{*}\right]_{\mu\nu}\left[\mvv{S}\mvv{O}-i\mvv{S}^{*}\mvv{O}^{*}\right]_{\mu\nu'} 
\int_{0}^{t} ds \int_{0}^{t} ds' \overline{\, \xi_{\nu}(s) \xi_{\nu'}(s') \,}\mbox{\hspace{0.6cm}}
\label{eterm1}
\\
\movel
 - \frac{i}{4} \sum_{\nu\nu'\lambda\lambda'}\left[\mvv{S}\mvv{O}+i\mvv{S}^{*}\mvv{O}^{*}\right]_{\lambda\nu}\left[\mvv{S}\mvv{O}+i\mvv{S}^{*}\mvv{O}^{*}\right]_{\lambda'\nu'} 
\int_{0}^{t} ds \int_{0}^{t} ds' \overline{\, n''_{\lambda}(s)n''_{\lambda'}(s')\xi_{\nu}(s) \xi_{\nu'}(s')\, }\mbox{\hspace{0.6cm}}
\label{eterm3}
\\
\movel
 - \frac{1}{2} \sum_{\nu\nu'\lambda'}\left[\mvv{S}\mvv{O}-i\mvv{S}^{*}\mvv{O}^{*}\right]_{\mu\nu}\left[\mvv{S}\mvv{O}+i\mvv{S}^{*}\mvv{O}^{*}\right]_{\lambda'\nu'} 
\int_{0}^{t} ds \int_{0}^{t} ds' \overline{\,n''_{\lambda'}(s')\xi_{\nu}(s) \xi_{\nu'}(s') \,}.\mbox{\hspace{0.6cm}}
\label{eterm2}
\end{eqnarray}
\end{subequations}
 
We consider the three terms in sequence:

\ssection{Term {\rm\bref{eterm1}}}
From independence of noises, $\overline{\xi_{\nu}(s)\xi_{\nu'}(s')} = \delta_{\nu\nu'}\,\delta(s-s')$ so that
\begin{equation}
\int_{0}^{t} ds \int_{0}^{t} ds' \overline{\, \xi_{\nu}(s) \xi_{\nu'}(s') \,}=t\,\delta_{\nu \nu'}.
\label{zz}
\end{equation}
 Hence
\begin{eqnarray}
\mbox{\eref{eterm1}}
&=& \frac{it}{4}\left[\mvv{S} \mvv{S}^{T}  -\left(\mvv{S} \mvv{S}^{T}\right)^{*} -i\mvv{S} \mvv{O} \mvv{O}^{\dagger} \mvv{S}^{\dagger} -i\left(\mvv{S} \mvv{O} \mvv{O}^{\dagger} \mvv{S}^{\dagger}\right)^* \right]_{\mu\mu}\nonumber\\
&=&\frac{t}{2}\left[\mvv{S} \mvv{O} \mvv{O}^{\dagger} \mvv{S}^{\dagger} \right]_{\mu\mu}.
\label{term1}
\end{eqnarray}
For the second line we note that $\mvv{S}\mvv{S}^T=\mvv{W}$ and $\mvv{W}$ is real, as well as that $(\mvv{S} \mvv{O} \mvv{O}^{\dagger} \mvv{S}^{\dagger})^*=(\mvv{S} \mvv{O} \mvv{O}^{\dagger} \mvv{S}^{\dagger})^T$. The transpose is then irrelevant for the $[\ ]_{\mu\mu}$ element.
Note that while $\mvv{O} \mvv{O}^{T} =\mvv{\id}$, this is not generally the case for $\mvv{O} \mvv{O}^{\dagger}$.

\ssection{Term {\rm\bref{eterm3}}}
The stochastic average is $\overline{\,n''_{\lambda}(s)n''_{\lambda'}(s')\xi_{\nu}(s) \xi_{\nu'}(s')\,}$. Now consider the following: if $s>s'$ then then $\xi(s)$ is uncorrelated with all the other terms, because of causality, and for the case of $n''(s)$, the non-anticipating nature of an It\^o process.  Hence, in this case we can write 
\begin{equation}
\movel
\overline{\,n''_{\lambda}(s)n''_{\lambda'}(s')\xi_{\nu}(s) \xi_{\nu'}(s')\,} = \overline{\,n''_{\lambda}(s)n''_{\lambda'}(s')\xi_{\nu'}(s')\,}\overline{\xi_{\nu}(s)} = 0\quad{\rm when}\quad s>s'\quad
\end{equation}
since $\overline{\xi_{\nu}(s)}=0$. 
The same argument can be made for $s'>s$, in which case a factor $\overline{\xi_{\nu'}(s')}=0$ occurs. 
The remaining situation is $s=s'$, in which case the noises $\xi$ are also uncorrelated with the $n''$ because of the non-anticipating nature. Hence, we can write
\begin{eqnarray}
\movel
\int_{0}^{t} ds \int_{0}^{s'} ds'\, \overline{\,n''_{\lambda}(s)n''_{\lambda'}(s')\xi_{\nu}(s) \xi_{\nu'}(s')\,}
\CR
= 
\int_{0}^{t} ds \int_{0}^{s'} ds'\,  \overline{\,n''_{\lambda}(s)n''_{\lambda'}(s')}\times \left\{
\begin{array}{c@{\quad{\rm when}\quad}l}\overline{\xi_{\nu}(s) \xi_{\nu'}(s')\,} & s=s'\\ 0 &  s\neq s'
\end{array}\right.
\CR
=
\int_{0}^{t} ds \int_{0}^{s'} ds'\, \overline{\,n''_{\lambda}(s)n''_{\lambda'}(s')}\delta(s-s')\delta_{\nu\nu'}
\CR
=
\delta_{\nu\nu'}\,\int_{0}^{t} ds\,\overline{\,n''_{\lambda}(s)n''_{\lambda'}(s)} = \delta_{\nu\nu'}{\cal N}''_{\lambda\lambda'}.
\label{term3averages}
\end{eqnarray}
The shorthand expression
\begin{equation}
Q = \int_0^t ds\, \overline{\,\bvv{n}''_{\lambda}(s)\bvv{n}''_{\lambda'}(s)^T}
\label{calN}
\end{equation}
for the correlation matrix of integrated expectation values will be useful. 
Inserting \eref{term3averages} into \eref{eterm3} gives
\begin{eqnarray}
\mbox{\eref{eterm3}}
&=& \frac{1}{4}\sum_{\lambda\lambda'}\left[\mvv{S} \mvv{O} \mvv{O}^{\dagger} \mvv{S}^{\dagger} +\left(\mvv{S} \mvv{O} \mvv{O}^{\dagger} \mvv{S}^{\dagger}\right)^* \right]_{\lambda\lambda'} Q_{\lambda\lambda'}
\CR
&=&\frac{1}{2}{\rm Tr}\left[{\rm Re}{\left[\mvv{S} \mvv{O} \mvv{O}^{\dagger} \mvv{S}^{\dagger} \right]} Q\right].
\label{term3}
\end{eqnarray}

\ssection{Term {\rm\bref{eterm2}}}
Proceeding in the same way, one finds
\begin{equation}
\movel
\overline{\,n''_{\lambda'}(s')\xi_{\nu}(s) \xi_{\nu'}(s')\,} =
\delta_{\nu\nu'}\,\int_{0}^{t} ds\,\overline{\,n''_{\lambda}(s)}.
\label{term2averages}
\end{equation}
and
\begin{eqnarray}
\mbox{\eref{eterm2}}
=\sum_{\lambda} \left( {\rm Im}\left[\mvv{S} \mvv{O} \mvv{O}^{\dagger} \mvv{S}^{\dagger}\right]_{\mu\lambda} -W_{\mu\lambda} \right)\,\int_{0}^{t}ds\, \overline{ n''_{\lambda}(s)}.
\label{term2}
\end{eqnarray}

Gathering now all the terms and inserting them into \eref{meanloglog} and \eref{charact_variance} we obtain:
\begin{eqnarray}
{\cal V}&=&{\cal V}(0) + \frac{1}{2M} \bigg\{
\frac{t}{2}\mbox{Tr}\left[\mvv{S} \mvv{O} \mvv{O}^{\dagger} \mvv{S}^{\dagger} \right]  
+M\mbox{Tr}\left[{\rm Re}\left\{\mvv{S} \mvv{O} \mvv{O}^{\dagger} \mvv{S}^{\dagger}\right\}  Q\right] 
\CR
&&+\sum_{\mu\lambda} \left( {\rm Im}\left[\mvv{S} \mvv{O} \mvv{O}^{\dagger} \mvv{S}^{\dagger}\right]_{\mu\lambda} -W_{\mu\lambda} \right)\,\int_{0}^{t}ds\, \overline{ n''_{\lambda}(s)}\,
\bigg\}.
\label{variance1}
\end{eqnarray}
%

\subsection{Stochastic density evolution} 

To proceed further we must evaluate the expectation values involving $\bvv{n}''$. Applying the It\^o formula for $n_m=\alpha_m\beta_m$, and again discarding the $\omega$ terms, leads to 
\begin{eqnarray}
\movel
\frac{dn_{m}}{dt}= \sqrt{i}\,n_{m}\sum_{\nu\sigma}\left[S_{m\nu}+S_{m+M,\nu}\right]O_{\nu\sigma}\xi_{\sigma} + in_m\sum_{\nu,\nu',\sigma} S_{m\nu} P_{\nu\sigma} S_{m+M,\nu'}O_{\nu'\sigma}.
\label{n_differential3}
\end{eqnarray}
Using the matrix $\mvv{F}$ defined in \eref{stratocorr_finalest}, this can be written
\begin{eqnarray}
\movel
\frac{dn_{m}}{dt} &=& \sqrt{i}\, n_m\left[ \mvv{F}\mvv{S}\mvv{O}\bvv{\xi}\right]_m + in_m \left[\mvv{W}(\mvv{F}-\mvv{I})\right]_{mm}
\CR
&=& \sqrt{i} n_m\left[ \mvv{F}\mvv{S}\mvv{O}\bvv{\xi}\right]_m.
\label{neqn}
\end{eqnarray}
The last follows because $\mvv{W}(\mvv{F}-\mvv{I})$ has all zeros on the diagonal. 
Moving to logarithmic variables and formally integrating the differential we obtain
\begin{eqnarray}
\movel
n_{\mu}(t)= n_{\mu}(0)\exp\left\{ \sqrt{i} [\mvv{F} \mvv{S} \mvv{O} \bv{\zeta}(t)]_{\mu}  \right\}.
\label{n_differential4}
\end{eqnarray}
where 
\begin{eqnarray}
\zeta_{\mu}(t)=\int_{0}^{t}\xi_{\mu}(s)ds.
\label{wiener}
\end{eqnarray}
Since it is a sum of Gaussian random variables, $\zeta_{\mu}$ is itself Gaussian distributed. It has correlations 
\begin{equation}
\overline{\zeta_{\mu}(t)\zeta_{\nu}(t')} = \delta_{\mu\nu}{\rm min}\left[t,t'\right],
\label{zetavar}
\end{equation}
as found in\cite{deuar:thesis}.

\subsection{Stochastic density means and variances} 

The expectation value of \eref{n_differential4} can be written in terms of the probability distribution of Gaussian random variables $\zeta_{\nu}$ that obey \eref{zetavar}:
\begin{eqnarray}
\movel
\overline{n_{\mu}(t)}
= \overline{n_{\mu}(0)}\int d^{2M}\bvv{\zeta}\,\mbox{Pr}[\bvv{\zeta},t]
\exp\left\{ \bvv{K}^{(\mu)}\cdot \bvv{\zeta}  \right\}. 
\label{nexpec1}
\end{eqnarray}
Here, $\bvv{K}^{(\mu)}$ are vectors with elements $K_{\nu}^{(\mu)}=\sqrt{i}\left[\mvv{F} \mvv{S} \mvv{O}\right]_{\mu\nu}$, and the probability distribution is 
\begin{eqnarray}
\mbox{Pr}[\bvv{\zeta},t]=\prod_{\sigma=1}^{2M}\frac{1}{\sqrt{2\pi t}}\exp\left[-\frac{\zeta_{\sigma}^{2}}{2t}\right]=\frac{1}{(2\pi t)^{M}}\exp\left[-\frac{\bvv{\zeta}\cdot \bvv{\zeta}}{2t}\right].
\label{multidimprob}
\end{eqnarray}
We can then write the expectation value as an integral over the quadratic form
\begin{eqnarray}
\overline{n_\mu(t)}= \frac{\overline{n_{\mu}(0)}}{(2\pi t)^M}\int d^{2M}\bvv{\zeta}
\exp\left\{-\bvv{\zeta}^{T}\mvv{A}(t)\bvv{\zeta}+ \bvv{K}^{(\mu)} \cdot \bvv{\zeta}  \right\},
\label{nexpec2}
\end{eqnarray}
where $\mvv{A}=\mvv{\id}/2t$, whence det$\mvv{A}=(2t)^{-2M}$. Using the standard result that
\begin{eqnarray}
\int d^{n}\bv{x}
\exp \left\{
-\bv{x}^{T}A\:\bv{x}+ \bv{K}\cdot\bv{x} 
\right\}
=\sqrt{  \frac{\pi^n}{\mbox{det}A}}  \exp\left(\frac{1}{4} \bv{K}^T A^{-1}\bv{K} \right),
\label{gaussianintegral}
\end{eqnarray}
the expectation value finally becomes
\begin{eqnarray}
\movel
\overline{n_\mu(t)}= \overline{n_{\mu}(0)}\exp\left\{\frac{t}{2}\bvv{K}^{(\mu)}\cdot \bvv{K}^{(\mu)}\right\} = \overline{n_{\mu}(0)}\exp\left\{\frac{it}{2}[\mvv{F}\mvv{W}\mvv{F}^{T}]_{\mu\mu}\right\}.
\label{nexpec3}
\end{eqnarray}
Note that for our form of matrices $W$ and $F$ in \eref{unified_eoms_posp} and \eref{stratocorr_finalest}, 
\begin{eqnarray}
\mvv{F}\mvv{W}\mvv{F}^{T}=\left( 
\begin{array}{cc}
0 & 0 \\
0 & 0
\end{array} 
\right).
\label{fwft}
\end{eqnarray}
Hence $\overline{n_\mu(t)}= \overline{n_{\mu}(0)}$ and 
\begin{equation}
\int_{0}^{t}ds\, \overline{ n''_{\lambda}(s)}=t\,\overline{n''_{\lambda}(0)}.
\label{intnii}
\end{equation}

We still have to evaluate $Q$ of \eref{calN}. Note that for $z,w\in{\mathbb C}$ we have $z''w''={\rm Re}(z w^{*} -zw )/2$. Thus
\begin{eqnarray}
\movel
Q_{\lambda\lambda'}= \overline{ \int_{0}^{t} n''_{\lambda}(s)n''_{\lambda'}(s)ds }
=\frac{1}{2}{\rm Re}\left\{ \int_{0}^{t} \left[\overline{ n_{\lambda}(s)n^{*}_{\lambda'}(s) } -\overline{ n_{\lambda}(s)n_{\lambda'}(s) }  \right]ds \right\} .
\label{nkkp1}
\end{eqnarray}
Using \eref{n_differential4} we can write:
\begin{eqnarray}
\movel
\overline{n_\lambda(t)n^{*}_{\lambda'}(t)}= \overline{n_{\lambda}(0)n^{*}_{\lambda'}(0)}\overline{\exp\left\{ \sqrt{i} [\mvv{F} \mvv{S} \mvv{O} \bvv{\zeta}(t)]_{\lambda} + \sqrt{-i} [\mvv{F} \mvv{S}^{*} \mvv{O}^{*} \bvv{\zeta}(t)]_{\lambda'}  \right\} }
\CR
= \overline{n_{\lambda}(0)n^{*}_{\lambda'}(0)}\int d^{2M}\bvv{\zeta}\,\mbox{Pr}[\bv{\zeta},t]
\exp\left\{ \bvv{K}^{(\lambda,\lambda')}\cdot \bvv{\zeta}  \right\}.
\label{nnstexpec1}
\end{eqnarray}
Now the symbol $\bvv{K}^{(\lambda,\lambda')}$ is 
\begin{eqnarray}
K_{\nu}^{(\lambda,\lambda')}=\sqrt{i}\left[\mvv{F}\mvv{S}\mvv{O}\right]_{\lambda\nu} +\sqrt{-i}\left[\mvv{F}\mvv{S}^{*}\mvv{O}^{*}\right]_{\lambda'\nu}.
\label{Bkkp}
\end{eqnarray}
To use \eref{gaussianintegral}, we obtain 
\begin{eqnarray}
\movel
\bvv{K}^{(\lambda,\lambda')}\cdot\bvv{K}^{(\lambda,\lambda')}
=i \left[\mvv{F}\mvv{W}\mvv{F} \right]_{\lambda\lambda}-i\left[\mvv{F}\mvv{W}\mvv{F} \right]_{\lambda'\lambda'} +2\left[\mvv{F} \mvv{S} \mvv{O}\mvv{O}^{\dagger} \mvv{S}^{\dagger} \mvv{F}\right]_{\lambda\lambda'}
\CR
=2 \left[\mvv{F} \mvv{S} \mvv{O}\mvv{O}^{\dagger}\mvv{S}^{\dagger}\mvv{F}\right]_{\lambda\lambda'}.
\label{Bkkp2}
\end{eqnarray}
We have used  $\mvv{F}\mvv{W}\mvv{F}^{T}=0$ and the other obvious properties of $\mvv{F}$, $\mvv{S}$, $\mvv{O}$.
Thus
\begin{eqnarray}
\overline{n_\lambda(s)n^{*}_{\lambda'}(s)}= \overline{n_{\lambda}(0)n^{*}_{\lambda'}(0)}\exp\left\{s \left[\mvv{F} \mvv{S} \mvv{O}\mvv{O}^{\dagger}\mvv{S}^{\dagger}\mvv{F}\right]_{\lambda\lambda'}\right\}.
\label{nnstexpec2}
\end{eqnarray}
Using the same methods we obtain
\begin{eqnarray}
\movel
\overline{n_\lambda(s)n_{\lambda'}(s)}&=& \overline{n_{\lambda}(0)n_{\lambda'}(0)}\exp\left\{\frac{is}{2} \left(\left[\mvv{F}\mvv{W}\mvv{F}\right]_{\lambda\lambda} 
+\left[\mvv{F}\mvv{W}\mvv{F}\right]_{\lambda'\lambda'} +2\left[\mvv{F}\mvv{W}\mvv{F}\right]_{\lambda\lambda'} \right) \right\}\CR
&=& \overline{n_{\lambda}(0)n_{\lambda'}(0)}.
\label{nnstexpec3}
\end{eqnarray}
Now we have to integrate over time. The result is \eref{Qresult}. 
We also make use of \eref{intnii} and the proprtties of $\mvv{W}$ and $\bvv{n}$ to find that 
\begin{equation}
\movel
\sum_{\mu\lambda}W_{\mu\lambda}\int_0^tds\,n''_{\lambda}(s)= t\sum_{\mu\lambda}W_{\mu\lambda} n''_{\lambda}(0) = \sum_{\mu}[\mvv{W}\bvv{n}]_{\mu} = \sum_m[-W\bv{n} + W\bv{n}]_{m} = 0.\quad
\label{Wn}
\end{equation}
Finally, 
substituting \eref{Wn} and \eref{Qresult}  into \eref{variance1}, we obtain the pivotal result \eref{variance_final}.

\section{Characteristic variance: without drift gauge}
\label{charvarappP}
The case when there is no drift gauge ($\bvv{f}=0$), 
follows in the same general way as above in appendix~\ref{charvarapp}, though with some additional elements. Summarising the procedure, we begin with
\begin{eqnarray}
\movel
\log{[|\gamma_{\mu}(t)|]}
&=&
\log{[|\gamma_{\mu}(0)|]} 
\CR
&&\hspace*{-2em}+ \frac{\sqrt{i}}{2} \sum_{\nu}\left[\mvv{S}\mvv{O} -i\mvv{S}^*\mvv{O}^{*}\right]_{\mu\nu}\int_{0}^{t}\xi_{\nu}(s)ds 
- \sum_{\nu}W_{\mu\nu} \int_{0}^{t}n''_{\nu}(s).\qquad
\label{logmodtP}
\end{eqnarray}
With the usual assumptions, $\bvv{n}(t)$ continues to be described by \eref{n_differential4}, and $\overline{n_{\mu}} = n_{\mu}(0)$. 
With the use of this, and algebra
\begin{eqnarray}
\movel
{\cal V}^{(P)}={\cal V}(0) + \frac{1}{2M} \bigg\{
\frac{t}{2}\mbox{Tr}\left[\mvv{S} \mvv{O} \mvv{O}^{\dagger} \mvv{S}^{\dagger} \right]  
+ {\rm Tr}\left[W Q^{(P)} W\right] -2t\sum_{\mu\lambda} W_{\mu\lambda}{\rm covar}\left[\log|\gamma_{\mu}(0)|,n''_{\lambda}(0)\right]
\CR
-\sqrt{i}\sum_{\mu\lambda\lambda'} \left[\mvv{S}\mvv{O}-i\mvv{S}^*\mvv{O}^*\right]_{\mu\lambda} W_{\lambda'\mu} \,\int_{0}^{t}ds\, \overline{\zeta_{\lambda}(t)n''_{\lambda'}(s)}
\bigg\},\qquad\quad
\label{variance1P}
\end{eqnarray}
in terms of a quantity
\begin{equation}
Q^{(P)}_{\lambda\lambda'} = \int_{0}^{t}ds\int_{0}^{t}ds'\, {\rm covar}\left[n''_{\lambda}(s),n''_{\lambda'}(s')\right]
\label{Qtdef}
\end{equation}
that is similar to $Q$.

To evaluate \eref{variance1P} and \eref{Qtdef} the two-time averages $\overline{n''_{\lambda}(s)n''_{\lambda'}(s')}$ and $\overline{\zeta_{\lambda}(t)n''_{\lambda'}(s)}$ are needed. From the construction \eref{wiener} we note that when $s'>s$, the random variable $\zeta$ can be split into completely independent early and late ($\widehat{\zeta}$) parts
\begin{equation}
\zeta_{\mu}(s') = \zeta_{\mu}(s) + \widehat{\zeta}_{\mu}(s'-s)
\label{zetasplit}
\end{equation}
with $\widehat{\zeta}(v)$ having the same variances as $\zeta(v)$ given in\eref{zetavar}, except that it is independent  because it consists of different building block noises $\xi(t)$. This way, after using \eref{n_differential4} and \eref{zetasplit} one has
\begin{eqnarray}
\movel\!\!\!\!\!\!
\int_{0}^{t}ds\int_{0}^{t}ds'\, \overline{n_{\lambda}(s)n^*_{\lambda'}(s')}  = 2\int_{0}^{t}ds\int_{s}^{t}ds'\, \overline{n_{\lambda}(s),n^*_{\lambda'}(s')}
\CR
= 2\int_{0}^{t}ds\int_{s}^{t}ds'\, \overline{n_{\lambda}(s),n^*_{\lambda'}(s)\exp\left\{ \sqrt{-i} [\mvv{F} \mvv{S}^* \mvv{O}^* \widehat{\bv{\zeta}}(s'-s)]_{\lambda'}  \right\}\,}\CR
= 2\int_{0}^{t}ds\,\overline{n_{\lambda}(s),n^*_{\lambda'}(s)}\int_{s}^{t}ds'\, \overline{\exp\left\{ \sqrt{-i} [\mvv{F} \mvv{S}^* \mvv{O}^* \widehat{\bv{\zeta}}(s'-s)]_{\lambda'}  \right\}\,}.\qquad\quad
\label{covarexpr1}
\end{eqnarray}
The factor in the right-hand integral can be identified as a scaled expectation value from \eref{nexpec3},  $\overline{n_{\lambda'}(s'-s)}^*/ \overline{n^*_{\lambda'}(0)}$, and substituting $s'=v+s$ one has
\begin{eqnarray}
\movel
\int_{0}^{t}ds\int_{0}^{t}ds'\,\overline{n_{\lambda}(s)n^*_{\lambda'}(s')}
=2\int_{0}^{t}ds\,\overline{n_{\lambda}(s),n^*_{\lambda'}(s)}\left[\int_{0}^{t-s}dv\, \frac{\overline{n_{\lambda'}(v)}}{\overline{n_{\lambda'}(0)}}\right]^*\CR
= 2\int_{0}^{t}ds\,\overline{n_{\lambda}(s),n^*_{\lambda'}(s)} (t-s)\CR
= 2\overline{n_{\lambda}(0)n^{*}_{\lambda'}(0)\int_{0}^{t}ds\, (t-s)}\exp\left\{s \left[\mvv{F} \mvv{S} \mvv{O}\mvv{O}^{\dagger}\mvv{S}^{\dagger}\mvv{F}\right]_{\lambda\lambda'}\right\}\CR
= \frac{2\overline{n_{\lambda}(0)n^{*}_{\lambda'}(0)}}{\left[\mvv{F} \mvv{S} \mvv{O}\mvv{O}^{\dagger}\mvv{S}^{\dagger}\mvv{F}\right]^2_{\lambda\lambda'}}
\left(e^{t\left[\mvv{F} \mvv{S} \mvv{O}\mvv{O}^{\dagger}\mvv{S}^{\dagger}\mvv{F}\right]_{\lambda\lambda'}} -1 -t\,\left[\mvv{F} \mvv{S} \mvv{O}\mvv{O}^{\dagger}\mvv{S}^{\dagger}\mvv{F}\right]_{\lambda\lambda'}
\right).\qquad\quad
\end{eqnarray}
Here,  \eref{intnii} was used in the 2nd line, and \eref{nnstexpec2} in the third. Similarly, 
\begin{equation}
\int_0^tds\,\int_0^tds'\,\overline{n_{\lambda}(s)n_{\lambda'}(s')} = t^2\,\overline{n_{\lambda}(0)n_{\lambda'}(0)}
\end{equation}
and $\int_0^tds\,\int_0^tds'\,\overline{n''_{\lambda}(s)}\overline{n''_{\lambda'}(s')} = t^2\,\overline{n''_{\lambda}(0)}\overline{n''_{\lambda'}(0)}$ from \eref{intnii}, so that using a form similar to \eref{nkkp1} one has 
\begin{eqnarray}
\movel
Q^{(P)}_{\lambda\lambda'} = 
{\rm Re}\bigg[\frac{\overline{n_{\lambda}(0)n^{*}_{\lambda'}(0)}}{\left[\mvv{F} \mvv{S} \mvv{O}\mvv{O}^{\dagger}\mvv{S}^{\dagger}\mvv{F}\right]^2_{\lambda\lambda'}}
\left(e^{t\left[\mvv{F} \mvv{S} \mvv{O}\mvv{O}^{\dagger}\mvv{S}^{\dagger}\mvv{F}\right]_{\lambda\lambda'}} -1 -t\,\left[\mvv{F} \mvv{S} \mvv{O}\mvv{O}^{\dagger}\mvv{S}^{\dagger}\mvv{F}\right]_{\lambda\lambda'}
\right)\CR
\quad - \frac{t^2}{2}\overline{n_{\lambda}(0)n_{\lambda'}(0)}
\bigg] 
- t^2\,\overline{n''_{\lambda}(0)}\overline{n''_{\lambda'}(0)}.\mbox{\hspace{3cm}}
\end{eqnarray}
This can be rearranged to the form 
\begin{eqnarray}
Q^{(P)}_{\lambda\lambda'} = \widetilde{Q}_{\lambda\lambda'} + t^2\,{\rm covar}\left[n''_{\lambda}(0),n''_{\lambda'}(0)\right]
\label{QP}
\end{eqnarray}
using $\widetilde{Q}$ from \eref{Qtilde}.
Consider now the quantity $\overline{\zeta_{\lambda}(t)n''_{\lambda'}(s)}$ in the last line of \eref{variance1P}. Splitting $\zeta$ into early and late parts as $\zeta(t) = \zeta(s)+\widehat{\zeta}(t-s)$ like in \eref{zetasplit}, one an see that $\widehat{\zeta}(t-s)$ must be uncorrelated with the earlier $n''(s)$. From this, and \eref{n_differential4} one has
\begin{equation}
\overline{\zeta_{\lambda}(t)n''_{\lambda'}(s)} = {\rm Im}\left\{ \overline{n_{\lambda'}(0)} \overline{\zeta_{\lambda}(s) e^{\sqrt{i}\left[\mvv{F}\mvv{S}\mvv{O}\bvv{\zeta}(s)\right]_{\lambda'}}}\right\}.
\end{equation}
The right-most expectation value is a Gaussian integral similar to \eref{nexpec2}, but the first moment of the random variable $\zeta(s)$. The matrix $\mvv{A}$ is now $\mvv{1}/(2s)$ and we have $\bvv{K}^{(\nu')}_\sigma = \sqrt{i}\left[\mvv{F}\mvv{S}\mvv{O}\right]_{\nu'\sigma}$.  
Here, the result from multivariate normal distributions that we need is 
\begin{eqnarray}
\movel
\int d^{n}\bv{x}\,
x_{\mu}\,
e^{
-\bv{x}^{T}A\:\bv{x}+ \bv{K}\cdot\bv{x} }
=\frac{1}{2}\left[\bv{K}^TA^{-1}\right]_{\mu}\sqrt{  \frac{\pi^n}{\mbox{det}A}}  \exp\left(\frac{1}{4} \bv{K}^T A^{-1}\bv{K} \right).
\label{gaussianintegral1}
\end{eqnarray}
With this one finds 
\begin{equation}
\overline{\zeta_{\lambda}(t)n''_{\lambda'}(s)} = s\,{\rm Im}\left\{\sqrt{i}\left[\mvv{F}\mvv{S}\mvv{O}\mvv{N}\right]_{\lambda'\lambda} \right\},
\end{equation}
where $\mvv{N} = {\rm diag}\left[\expec{\bvv{n}(0)}\right]$ as defined before. The bottom line in \eref{variance1P} is
\begin{equation}
-\frac{t^2}{2}{\rm Tr}\left\{ 
\mvv{W}\mvv{N}'\mvv{F}\mvv{W} - {\rm Im}\left[ \mvv{S}\mvv{O}\mvv{O}^{\dagger}\mvv{S}^{\dagger}\mvv{N}^*\right] \mvv{F}\mvv{W} 
\right\}.
\end{equation}
Finally substituting this and \eref{QP}, the expression \eref{variance_finalP} is obtained after some manipulation to separate the mostly useless initial covariance $C^{(0)}$ defined in \eref{C0}.

\section{Noise performance improvement for large systems}
\label{noiserewriting}
As described in \sref{pp_altnoise}, a more efficient formulation of the Gauge-P SDEs is possible using a noise field $\chi_{\wt{p}}$ with correlation properties 
\begin{equation}\label{chichi}
\overline{ \chi_{\wt{p}} \chi_{\wt{q}} } = \delta_{\wt{p},-\wt{q}}.
\end{equation}
Such a noise can indeed be constructed, by first dividing the k-space of the problem into three regions: $\mathcal{R}$ contains half of all the k-space vectors $\mv{k}$ that possess a matching wavevector $-\mv{k}$ reflected across the origin. In 3D on a standard DFT lattice $\mathcal{R}$ can consist for example of all wavevectors in the $k_x>0$ half-space, plus all vectors on the $k_x=0$ plane with $k_y>0$, plus all vectors on the $k_z$ axis with $k_z>0$. $\mathcal{R}'$ contains all the negatives $-\mv{k}$ of the wavevectors $\mv{k}\in\mathcal{R}$. Finally, $\mathcal{R}_0$ contains the residual unpaired wavevectors. We always have $[\mv{k}=0] \in \mathcal{R}_0$, but if the number of lattice points $M_d$ in a particular dimension $d$ is even, it will also contain the unpaired wavevectors possessing maximally negative values $k_d=-\pi L_d/M_d$ that occur in the DFT for box length $L_d$.  With this division, the prescription for \eref{chichi} is
\begin{equation}\label{chi}
\chi_{\wt{m}} = \left\{\begin{array}{c@{\quad{\rm if}\quad}l}
\frac{1}{\sqrt{2}}\left[\xi_{(\wt{m},1)} + i\xi_{(\wt{m},2)}\right] & \mv{k}_{\wt{m}}\in\mathcal{R}\\
\frac{1}{\sqrt{2}}\left[\xi_{(-\wt{m},1)} - i\xi_{(-\wt{m},2)}\right] & \mv{k}_{\wt{m}}\in\mathcal{R}'\\
\xi_{(\wt{m},0)} &\mv{k}_{\wt{m}}\in\mathcal{R}_0
\end{array}\right. .
\end{equation}
The real noises $\xi$ have the same properties as before, i.e., they are all independent with variance $\overline{\xi_{(\wt{p},j)}(t)\xi_{(\wt{q},j')}(t')} = \delta_{\wt{p}\wt{q}}\,\delta_{jj'}\,\delta(t-t')$.

\section{Non-local gauge expressions}
\label{nonlocalDG}

The purpose of this appendix is to elaborate how the results of \sref{analytical_nonlocal_diff} were obtained. We desire a simpler version of 
\eref{variance_final} valid for short times. To this end, the exponential within each element $\lambda,\lambda'$ of the matrix $\left(e^{t\left[\mvv{F} \mvv{S} \mvv{O}\mvv{O}^{\dagger}\mvv{S}^{\dagger}\mvv{F}\right]_{\lambda\lambda'}} -1 \right)$ in \eref{variance_final} is expanded to second order in $t$. One obtains
\begin{eqnarray}
\movel
{\cal V}=\frac{t}{2M}\mbox{Tr}\left[\cosh(2 A) W \right]  + t\,\bv{n}^{\prime\prime T}(0)\sqrt{{W}} \exp(-2 A)\sqrt{{W}} \bv{n}''(0)
\CR
+ \frac{t^2}{2}\mbox{Tr}\left[\sqrt{{W}} \exp(-2 A)\sqrt{{W}} {N}' \sqrt{{W}} \exp(-2 A)\sqrt{{W}} {N}' + ({N}' \rightarrow {N}'' )\right], 
\label{variance_nonlocalO_app}
\end{eqnarray}
as written in \eref{variance_nonlocalO}, with the diagonal matrix $N$ given by \eref{NiNii}.

The more difficult task, is to the differentiate \eref{variance_nonlocalO_app} with respect to each element $a_{ij}$ of the diffusion gauge matrix $A$. We make use of the expression for derivatives of matrix exponentials:
\begin{eqnarray}
\frac{\partial \exp{[A]}  }{\partial a_{ij}  }=\int_{0}^1d\tau \exp{[(1-\tau)A]}\frac{\partial A}{\partial a_{ij}}\exp{[\tau A]},
\end{eqnarray}
where $[\partial A /\partial a_{ij}]_{lm}=\delta_{li}\delta_{mj}$. With this expression we require:
\begin{eqnarray}
\movel
0=\frac{\partial{\cal V}  }{\partial a_{ij}  }
=\int_{0}^1d\tau\bigg\{
\frac{t}{2M}\big\{ \left[\exp{[2 \tau A]} W \exp{[2 (1-\tau) A]} \right]_{ji}
\CR
\movel
-\left[\exp{[-2 \tau A]} W \exp{[-2 (1-\tau) A]} \right]_{ji}\big\}
-2 t\left[\exp{[-2 \tau A]} \sqrt{W}{\bv{n}''(0)\bv{n}^{\prime\prime T}(0)}\sqrt{W}\exp{[-2 (1-\tau) A]} \right]_{ji}
\CR
\movel
-2 t^2 \left[\exp{[2 \tau A]} \sqrt{W}N' \sqrt{W}e^{-2A}\sqrt{W}N'\sqrt{W} \exp{[2 (1-\tau) A]} + \{ N' \rightarrow N''  \}\right]_{ji}
\bigg\}.\mbox{\hspace{2cm}}
\label{variance_derivative_bare}
\end{eqnarray}
Let us assume that $A$ will be a function of $\sqrt{W}$ only\footnote{Defined by a power series.}, and the initial mode occupation is constant and real $n_{0}(x)=n_{0}$. Now matrices such as $\exp{[2 \tau A]}$, $N'$ and $\sqrt{W}$ commute. The $\tau$ integration is then trivial, and we arrive at the matrix equation:
\begin{eqnarray}
W=W\exp(-4A) + 4M t \left[\sqrt{W}N'WN' \sqrt{W}\right] \exp(-6A), 
\end{eqnarray}
from which we can directly write \eref{nonlocalAmatrixequation} and then \eref{optimalA}.

\end{document}